\documentclass[10pt, journal]{IEEEtran}
\usepackage{amsthm}
\usepackage{amsmath}
\usepackage{graphicx}
\usepackage{amssymb}
\usepackage{epstopdf}
\usepackage{xargs}
\usepackage{ifthen}
\usepackage{dsfont}
\usepackage{aliascnt}
\usepackage{stmaryrd}
\usepackage{mathtools}
\mathtoolsset{showonlyrefs}
\usepackage{enumerate}
\usepackage{ushort}
\usepackage{algpseudocode}
\usepackage{algorithm}
\usepackage{tikz}
\usetikzlibrary{positioning}
\usetikzlibrary{patterns}
\usepackage{hyperref}
\usepackage{color}

\newtheorem{theorem}{Theorem}
\newaliascnt{proposition}{theorem}

\aliascntresetthe{proposition}

\newaliascnt{lemma}{theorem}
\newtheorem{lemma}[lemma]{Lemma}
\aliascntresetthe{lemma}

\newaliascnt{corollary}{theorem}

\aliascntresetthe{corollary}

\newaliascnt{definition}{theorem}

\aliascntresetthe{definition}

\newaliascnt{example}{theorem}

\aliascntresetthe{example}

\newaliascnt{remark}{theorem}
\newtheorem{remark}[remark]{Remark}
\aliascntresetthe{remark}

\newtheorem{assumption}{Assumption}

\newcounter{hypA}

\newcounter{hypB}

\def\equationautorefname~#1\null{%
  Equation~(#1)\null
}

\newcommand{\1}[1]{\mathds{1}_{#1}}

\newcommandx\A[2][1=]{
\ifthenelse{\equal{#1}{}}
{\hspace{-1mm}(\textbf{A\ref{#2}})\hspace{-1mm}}
{\hspace{-1mm}(\textbf{A\ref{#1}--\ref{#2}})\hspace{-1mm}}
}
\newcommand{\act}[1]{\mathfrak{A}_{#1}}
\newcommand{\af}[1]{\termletter_{#1}}
\newcommand{\alg}[1]{\mathcal{#1}} 

\newcommand{\addf}[1]{\tilde{h}_{#1}} 
\newcommandx{\arr}[2][1=]{
\ifthenelse{\equal{#1}{}}
{\upsilon_{\N}^{#2}}
{(\upsilon_{\N}^{#2})^{#1}}
}
\newcommandx{\arrterm}[3][1=]{
\ifthenelse{\equal{#1}{}}
{\tilde{\upsilon}_{\N}(#3,#2)}
{\tilde{\upsilon}_{\N}^{#1}(#3,#2)}
}
\newcommandx{\asvar}[4][1=]{
\ifthenelse{\equal{#1}{}}
{\sigma_{#2} \langle #3, #4 \rangle(\af{#1})}
{\sigma_{#2}^2 \langle #3, #4 \rangle(\af{#2})}
}
\newcommandx{\asvarFFBSm}[4][1=]{
\ifthenelse{\equal{#1}{}}
{\tilde{\sigma}_{#2} \langle #3, #4 \rangle(\af{#1})}
{\tilde{\sigma}_{#2}^2 \langle #3, #4 \rangle(\af{#2})}
}
\newcommandx{\asvarstd}[2][1=]{
\ifthenelse{\equal{#1}{}}
{\sigma_{#2}(\af{#2})}
{\sigma_{#2}^2(\af{#2})}
}
\newcommandx{\asvarFFBSmstd}[2][1=]{
\ifthenelse{\equal{#1}{}}
{\tilde{\sigma}_{#2}(\af{#2})}
{\tilde{\sigma}_{#2}^2(\af{#2})}
}

\newcommandx\B[2][1=]{
\ifthenelse{\equal{#1}{}}
{\hspace{-1mm}(\textbf{B\ref{#2}})\hspace{-1mm}}
{\hspace{-1mm}(\textbf{B\ref{#1}--\ref{#2}})\hspace{-1mm}}
}
\newcommandx{\BF}[3][1=]{
\ifthenelse{\equal{#1}{}}
{\kernel{D}_{#2, #3}}
{\kernel{D}_{#2, #3}^{#1}}
}
\newcommandx{\BFcent}[3][1=]{
\ifthenelse{\equal{#1}{}}
{\tilde{\kernel{D}}_{#2, #3}}
{\tilde{\kernel{D}}_{#2, #3}^{#1}}
}
\newcommand{\bi}[3]{J_{#1}^{(#2, #3)}}
\newcommandx{\bk}[2][1=]{ 
\ifthenelse{\equal{#1}{}}
{\overleftarrow{\kernel{Q}}_{#2}}
{\overleftarrow{\kernel{Q}}_{#2}^{#1}}
}
\newcommandx{\bd}[2][1=]{ 
\ifthenelse{\equal{#1}{}}
{\overleftarrow{\hd}_{\hspace{-.1cm}#2}}
{\overleftarrow{\hd}_{\hspace{-.1cm}#2}^{#1}}
}

\newcommandx{\cexp}[3][1=]{
\ifthenelse{\equal{#1}{}}
{\mathbb{E}\left[ #2 \mid #3 \right]} 
{\mathbb{E}[ #2 \mid #3 ]}
}

\newcommand{\convp}{\overset{\prob}{\longrightarrow}}

\newcommand{\E}{\mathbb{E}}
\newcommand{\epart}[2]{\xi_{#1}^{#2}}
\newcommand{\eqdef}{\vcentcolon=} 

\newcommandx{\genfd}[1][1=]{
\ifthenelse{\equal{#1}{}}
{\mathcal{F}}
{\mathcal{F}_{\N}}
}

\newcommand{\hbdnt}{|h|_{\infty}}
\newcommand{\hk}{\kernel{Q}} 
\newcommand{\hklow}{\ushort{\varepsilon}}
\newcommand{\hkup}{\bar{\varepsilon}}
\newcommand{\hd}{q} 

\renewcommand{\k}{j}
\newcommand{\kernel}[1]{\mathbf{#1}}
\newcommand{\kletter}{\tilde{\N}}
\newcommandx{\K}[1][1=]{
\ifthenelse{\equal{#1}{}}{{\kletter}}{{\tilde{\N}^{#1}}}
}

\newcommand{\lag}[1]{s_{\Delta}(#1)}

\newcommandx{\lebfun}[1][1=]{
\ifthenelse{\equal{#1}{}}
{\lebfunletter}
{\lebfunletter_{#1}}
}

\newcommand{\lebfunletter}{\varphi}
\newcommand{\lk}[1]{\kernel{L}_{#1}} 

\newcommandx{\M}[1][1=]{
\ifthenelse{\equal{#1}{}}
{\N}
{\N}
}
\newcommand{\md}[1]{g_{#1}} 
\newcommand{\mdlow}{\ushort{\delta}}
\newcommand{\mdup}{\bar{\delta}}

\newcommand{\measSpace}[1]{(\set{#1},\alg{#1})} 
\newcommand{\mr}{\varrho}

\newcommand{\mk}{\kernel{G}}

\newcommand{\N}{N}

\newcommand{\nset}{\mathbb{N}}
\newcommand{\nsetpos}{\mathbb{N}^*}

\newcommand{\ordo}{\mathcal{O}}
\newcommandx{\oscn}[2][1=]{
\ifthenelse{\equal{#1}{}}{\operatorname{osc}(#2)}{\operatorname{osc}^{#1}(#2)}
}

\newcommand{\PF}{\mathsf{PF}}
\newcommandx\post[2][1=]{
\ifthenelse{\equal{#1}{}}
	{\phi_{#2}}
	{\phi_{#2}^\N}
}

\newcommandx\postafl[2][1=]{
\ifthenelse{\equal{#1}{}}
	{\phi_{#2}^{\tol}}
	{\phi_{#2}^{\N,\tol}}
}
\newcommandx\posthat[2][1=]{
\ifthenelse{\equal{#1}{}}
	{\hat{\phi}_{#2}}
	{\hat{\phi}_{#2}^{\N}}
}

\newcommand{\prob}{\mathbb{P}} 
\newcommand{\probdist}{\mathsf{Pr}}

\newcommand{\probSpace}{(\Omega,\alg{F},\prob)} 

\newcommand{\refM}{\mu} 
\newcommand{\rmd}{\mathrm{d}}

\newcommand{\set}[1]{\mathsf{#1}} 

\newcommand{\sqc}[1]{\eta_{#1}}
\newcommand{\supn}[1]{\|#1\|_{\infty}} 

\newcommand{\termletter}{{h}}
\newcommand{\testfsymb}{f}
\newcommandx{\testf}[1][1=]{  
\ifthenelse{\equal{#1}{}}{\testfsymb}{\testfsymb_{#1}}
}
\newcommand{\testfpsymb}{\tilde{f}}
\newcommandx{\testfp}[1][1=]{  
\ifthenelse{\equal{#1}{}}{\testfpsymb}{\testfpsymb_{#1}}
}

\newcommand{\tol}{\varepsilon}

\newcommand{\tstatletter}{\kernel{T}}
\newcommandx\tstat[2][1=]{
\ifthenelse{\equal{#1}{}}
	{\tstatletter_{#2}}
	{\tau_{#2}^{#1}}
}
\newcommand{\tstattil}[2]{\tilde{\tau}_{#2}^{#1}}
\newcommand{\tstatfun}[1]{T_{#1}}

\newcommandx\varlim[2][1=]{
\ifthenelse{\equal{#1}{}}
	{\sigma_{#2}^{2, \infty}}
	{\sigma_{#2}^{2, \N}}
}

\newcommand{\wgt}[2]{\omega_{#1}^{#2}}
\newcommand{\wgtsum}[1]{\Omega_{#1}}

\newcommand{\Xinit}{\chi}

\begin{document}

	
\title{Particle-based adaptive-lag online marginal smoothing in general state-space models}

\author{Johan~Alenl\"ov
		and~Jimmy~Olsson%
\thanks{J. Alenl\"ov is with the Department of Information Technology, Uppsala University, Sweden, e-mail: johan.alenlov@it.uu.se}%
\thanks{J. Olsson is with the Department of Mathematics, KTH Royal Institute of Technology, Sweden, e-mail: jimmyol@kth.se}}%




\maketitle

\begin{abstract}
We present a novel algorithm, an adaptive-lag smoother, approximating efficiently, in an online fashion, sequences of expectations under the marginal smoothing distributions in general state-space models. The algorithm evolves recursively a bank of estimators, one for each marginal, in resemblance with the so-called particle-based, rapid incremental smoother (PaRIS). Each estimator is propagated until a stopping criterion, measuring the fluctuations of the estimates, is met. The presented algorithm is furnished with theoretical results describing its asymptotic limit and memory usage.  




\end{abstract}



\section{Introduction}
\label{sec:intro}

\IEEEPARstart{S}{tate-space} models (SSMs), also known as \emph{hidden Markov models} (HMMs), are fundamental in many scientific and engineering disciplines. Incorporating unobservable, Markovian states, these models are adjustable enough to model a variety of complex, real-world time series in, e.g., econometrics~\cite{chib:02}, speech recognition~\cite{rabiner:juang:1993}, and target tracking~\cite{mihaylova:angelova:honary:bull:canagarajah:ristic:2007}. In this paper we focus on online state reconstruction in SSMs; more specifically, our goal is to estimate, on-the-fly as new data appear, expectations under the marginal posteriors of the different states given the data. 

More precisely, an SSM is a bivariate model consisting of an observable process $\{Y_t\}_{t \in \nset}$, referred to as the \emph{observation process}, and an unobservable Markov chain $\{X_t\}_{t \in \nset}$, known as the \emph{state process}, taking on values in some general measurable spaces $\measSpace{Y}$ and $\measSpace{X}$, respectively. Throughout the paper the subindex $t$ will often be referred to as ``time'' without being necessarily a temporal index. 

When operating on SSMs one is often interested in calculating expectations under the conditional distribution of one or several states conditioned upon a subset of some given stream $\{ y_t \}_{t \in \nset}$ of observations. For any $(s,s',t) \in \nset^3$ such that $s \leq s' \leq t$ we denote by $\post{s:s' \mid t}$ the conditional distribution of $X_{s:s'} = (X_s, \ldots, X_{s'})$ (our notation for vectors) given $Y_{0:t} = y_{0:t}$ (a precise definition is given in Section~\ref{sec:prel}). Some special distributions of interest are the \emph{filter distributions} $\post{t} \eqdef \post{t:t\mid t}$, the \emph{joint smoothing distributions} $\post{0:t\mid t}$, and, finally, when $s = s'$, the \emph{marginal smoothing distributions} $\post{s \mid t}$.

For any probability measure $\mu$ and real-valued measurable function $h$, let $\mu h \eqdef \int h(x) \, \mu(\rmd x)$ denote the Lebesgue integral of $h$ with respect to $\mu$ (whenever this is well defined). Given an observation stream $\{ y_t \}_{t \in \nset}$ and a sequence $\{ \af{t} \}_{t \in \nset}$ of real-valued functions, the present paper focuses on online calculation of the flow 
\begin{equation} \label{eq:tar}
(\post{0 \mid t} \af{0}, \post{1 \mid t} \af{1}, \ldots, \post{s \mid t} \af{s}, \ldots, \post{t - 1 \mid t} \af{t - 1}, \post{t} \af{t}), \quad t \in \nset, 
\end{equation}
as $t$ increases. Note that a significant challenge with this problem is that \emph{all} elements of the vector \eqref{eq:tar} change with $t$. Moreover, in the problem formulation, the word ``online'' means, first, that the observation sequence is processed in a single sweep as new observations appear and, second, that the computational cost of updating the approximations one step, i.e., from $t$ to $t + 1$, is uniformly bounded in $t$. Hence, the aggregated computational cost and memory requirements of the algorithm should grow at most linearly with $t$. 

\begin{remark}
In the literature (see, e.g., \cite{briers:doucet:maskell:2010}), the problem of computing $\post{s \mid t}$ offline for a fixed $t$ and all $s \leq t$ is typically referred to as \emph{fixed-interval smoothing}. On the other hand, the problem of computing $\post{s \mid t}$ online for a fixed $s$ and increasing $t$ is typically referred to as \emph{fixed-point smoothing}. Our aim of computing the whole vector \eqref{eq:tar} online could hence be viewed as a combination of these two problems. 
\end{remark}

\subsection{Previous work}

Exact smoothing is possible only in models with finite state space and for linear Gaussian SSMs using the \emph{forward-backward smoother}~\cite{rabiner:1989} and the \emph{disturbance smoother}~\cite{dejong:1988} (sometimes referenced to as the \emph{Kalman smoother}), respectively. When dealing with general, possibly nonlinear SSMs, current methods take mainly two different approaches. The first approach relies, as in the \emph{extended Kalman filter} and the \emph{unscented Kalman filter}~\cite{anderson:moore:1979,julier:uhlmann:1997}, on linearisation of the model and Kalman filtering. These methods work well if the model is almost linear Gaussian, but will introduce significant errors in the presence of highly nonlinear model components. 

The second approach relies on Monte Carlo simulation, preferably in the form of \emph{sequential Monte Carlo} (SMC) \emph{methods}, or, \emph{particle filters}. Particle filters propagate recursively a sample of random simulations, so-called \emph{particles}, mimicking typically, as in the \emph{bootstrap particle filter}, the latent state (or, prior) dynamics. At each time step, the particles are associated with importance weights compensating for the discrepancy between the prior dynamics and the conditional (posterior) dynamics of the states given the observations. By duplicating and killing, through resampling, particles with high and low importance, respectively, the particles are directed towards regions of the state space with high posterior probability. 

At each time step, the weighted empirical measure associated with the particles serves as an approximation of the filter distribution at the time step in question. Moreover, it can be shown that the empirical measure formed by the ancestral lines of the particles provides an approximation also of the joint smoothing distribution. However, since the particles are resampled repeatedly, these trajectories collapse eventually, and in the long run such a Monte Carlo approximation will rely on more or less a single path. Thus, the statistician is referred to alternative techniques such as the \emph{forward-filtering backward-smoothing} (FFBSm)~\cite{doucet:godsill:andrieu:2000}, \emph{forward-filtering backward-simulation} (FFBSi)~\cite{godsill:doucet:west:2004} or \emph{joint backward simulation Rao-Blackwellized particle filter}~\cite{lindsten:bunch:godsill:schon:2013} algorithms, which approximate the so-called \emph{backward decomposition} of the joint smoothing distribution. The \emph{two-filter algorithm}~\cite{kitagawa:1996,briers:doucet:maskell:2010,fearnhead:wyncoll:tawn:2010,nguyen:lecorff:moulines:2017} combines two separate particle filters, one in the forward direction and one in the reverse direction. In the \emph{adaptive path integral smoother}~\cite{ruiz:kappen:2017}, smoothing is performed by solving a control problem. A popular class of methods are the \emph{Markov chain Monte Carlo} (MCMC) \emph{methods} simulating a Markov chain admitting the joint smoothing distribution as invariant distribution. Recently, \cite{andrieu:doucet:holenstein:2010} combined successfully particle methods and MCMC into \emph{particle Markov chain Monte Carlo} (PMCMC) \emph{methods} to construct such Markov chains.  All of these methods, in their most basic forms, require more than one pass of the data, either a forward and backward pass or that the estimates needs to be iterated until convergence is reached. Thus, the online processing setting of the present paper invalidates these methods.

As known to us, there is only one existing approach that applies to the problem \eqref{eq:tar} without violating the online criteria set up by us, namely the particle-based \emph{fixed-lag smoother}~\cite{kitagawa:sato:2001,olsson:cappe:douc:moulines:2006}. This is a genealogical tracing-based particle method that copes with particle path degeneracy by means of truncation. The truncation implies a bias that may be controlled using forgetting arguments (see \cite{olsson:cappe:douc:moulines:2006} for an analysis). A problem with this method, which will be explained in some detail in~\autoref{ssec:fls}, is that the truncation lag is a design parameter that needs to be set a priori by the user. This is nontrivial, as the optimal choice of the lag depends on the mixing properties of the model.

\subsection{Our Contribution}

In the present paper we introduce a novel algorithm that provides an approximate solution to the research question outlined in the previous section. The algorithm relies on a recursive form of the backward decomposition that has been employed previously for forward smoothing of additive state functionals~\cite{cappe:2009,delmoral:doucet:singh:2009}. The same decomposition is used in the \emph{particle-based, rapid incremental smoother} (PaRIS), proposed by the authors in~\cite{olsson:westerborn:2014} and analysed theoretically in~\cite{olsson:westerborn:2014b}, which performs online smoothing of additive state functionals with constant memory and computational complexity demands that grow only linearly with the number of particles (compared to the quadratic growth of other algorithms \cite{delmoral:doucet:singh:2009}). 

The method proposed in the present paper is driven by the same sampling technique as the PaRIS and could be likened to computing one PaRIS estimate per marginal in \eqref{eq:tar}. All these PaRIS estimates are formed on the basis of the output of the same underlying particle filter. However, since our aim is to compute, on-the-fly, \emph{all} the marginal expectations in \eqref{eq:tar}, we need, in order to avoid computational overload, to stop updating the estimate of a given marginal expectation once some criterion tells us to do so. This leads to an \emph{adaptive-lag approach}. A sensible such stopping criterion will be designed using the forgetting properties of the model.

Our contribution is presented in two steps: first, we consider an ideal algorithm, applicable in the context of linear Gaussian models and requiring closed-form computation of all quantities of interest; after this, the general---possible nonlinear/non-Gaussian---case is dealt with using particle-based approximations. By using results derived in \cite{olsson:westerborn:2014b}, we are able to analyse theoretically the asymptotic properties and memory demands of our algorithm in order to place the same on more solid ground.

Finally, we illustrate numerically the performance of the proposed algorithm and compare the same to that of the marginal fixed-lag smoother. As shown by the simulations, our adaptive-lag approach avoids completely the complex bias-variance tradeoff that is an unavoidable ingredient of optimal fixed-lag design. Moreover, it is shown that the marginal adaptive-lag smoother is on par with the optimally tuned fixed-lag smoother in terms of variance.  

\subsection{Outline}

In \autoref{sec:prel} we introduce formally SSMs and the backward decomposition. \autoref{sec:algo} introduces the ideal algorithm, providing a conceptual understanding of our approach without involving particle approximations. In~\autoref{sec:pf} we turn to the nonlinear/non-Gaussian case and replace intractable quantities by particle-based estimates. Theoretical results are presented in~\autoref{sec:theory} and in~\autoref{sec:simul} the algorithm is benchmarked numerically on two models. Our conclusions are presented in~\autoref{sec:conc} and, finally, \autoref{sec:append} contains some proofs.

\section{Preliminaries}
\label{sec:prel}
In the following, let $\nset$ and $\nsetpos$ denote the sets of nonnegative and positive natural numbers, respectively. For any bounded measurable function $h$ we let $\supn{h} \eqdef \sup_{x \in \set{X}}|h(x)|$ and $\oscn{h} \eqdef \sup_{(x,x') \in \set{X}^2}|h(x) - h(x')|$ denote the supremum and oscillator norms of $h$, respectively. 

In the following we assume that all random elements are well-defined on a common probability space $\probSpace{}$. Let $\measSpace{X}$ and $\measSpace{Y}$ be some measurable spaces, $\hk : \set{X} \times \alg{X} \to [0,1]$ and $\mk{} : \set{X} \times \alg{Y} \to [0,1]$ some Markov transition kernels, and $\Xinit{}$ a probability measure on $\alg{X}$. 
An SSM is a bivariate Markov chain $\{ (X_t, Y_t) \}_{t \in \nset}$ on $\set{X} \times \set{Y}$ such that $
X_{t + 1},Y_{t + 1} \mid X_t, Y_t \sim \hk(X_t, \rmd x_{t + 1}) \, \mk(x_{t + 1}, \rmd y_{t + 1})$ and $X_0, Y_0 \sim \Xinit{}(\rmd x_0) \, \mk(x_0, \rmd y_0)$. It is assumed that only $\{ Y_t \}_{t \in \nset}$ is observable. Using this definition, it is easily shown that 
\begin{enumerate}[(i)]
	\item the unobservable state sequence $\{X_t\}_{t \in \nset}$ is a Markov chain with transition kernel $\hk$ and initial distribution $\Xinit{}$.
	\item the variables of the observable process $\{Y_t\}_{t \in \nset}$ are, conditionally on the states, independent and such that the conditional distribution of each $Y_t$ depends on the corresponding $X_t$ only. 
\end{enumerate}

Throughout this paper we will assume that the model is \emph{fully dominated}, i.e., that $\hk$ and $\mk$ admit transition densities $\hd$ and $\md{}$, respectively, with respect to some reference measures. We will in the following assume that we are given a fixed stream $\{ y_t \}_{t \in \nset}$ of observations. For ease of notation, let for all $t \in \nset$, $\md{t}(x) \eqdef \md{}(x, y_t)$, $x \in \set{X}$. The joint smoothing distribution at time $t$, i.e., the law of $X_{0:t}$ conditionally to $Y_{0:t} = y_{0:t}$, is 
$$
\post{0:t \mid t}(\rmd x_{0:t}) \eqdef \frac{\Xinit{}(\rmd x_0) \md{0}(x_0) \prod_{s = 1}^t \hk(x_{s - 1}, \rmd x_s) \md{s}(x_s)}{\idotsint \Xinit{}(\rmd x'_0) \md{0}(x'_0) \prod_{s = 1}^t \hk(x'_{s - 1}, \rmd x'_s) \md{s}(x'_s)}, 
$$
and all posteriors $\post{s:s' \mid t}$ of interest (including the filter $\post{t}$) are obtained as marginals of $\post{0:t \mid t}$. 

Interestingly, the state process is still Markov when evolving conditionally to $Y_{0:t}$ in the time-reversed direction; in particular, using Bayes' formula and the Markov property of the bivariate process $\{ (X_t, Y_t) \}_{t \in \mathbb{N}}$ it is checked straightforwardly that the distribution of $X_s$ given $X_{s+1} = x_{s + 1}$ and $Y_{0:s} = y_{0:s}$ is given by 
\begin{equation} 
	\bk{\post{s}}(x_{s+1}, \rmd x_s) \eqdef \frac{\hd(x_s, x_{s+1}) \post{s}(\rmd x_s)}{\int \hd(x'_s, x_{s+1}) \post{s}(\rmd x'_s)}, \quad s \in \nset  \label{eq:bk}
\end{equation}
(see, e.g.,~\cite[Prop.~3.3.6]{cappe:moulines:ryden:2005}). 
Using \eqref{eq:bk}, the joint-smoothing distribution may be expressed by the \emph{backward decomposition}
$$
    \post{0:t \mid t}(\rmd x_{0:t}) = \post{t}(\rmd x_t) \prod_{s = 0}^{t - 1} \bk{\post{s}}(x_{s+1}, \rmd x_s), 
$$ 
which is instrumental in many smoothing procedures~\cite{godsill:doucet:west:2004,douc:garivier:moulines:olsson:2010,olsson:westerborn:2014b,delmoral:doucet:singh:2009}.

\section{Online marginal smoothing}
\label{sec:algo}

Recall that our aim is to estimate the vectors \eqref{eq:tar} in an online manner as $t$ increases. For the moment, consider estimation of some expectation $\post{s \mid t} \af{s}$ under some marginal $\phi_{s \mid t}$. Under suitable ergodicity conditions (to be specified later), we may expect observations of the distant future to have limited effect on the posterior of some state $X_s$. Consequently, we may expect $\post{s \mid t} \af{s}$ to converge to some fixed point $\post{s \mid \infty} \af{s}$ as $t$ increases; see~\cite[Sec. 4.3]{cappe:moulines:ryden:2005}. Thus, allowing for a negligible bias, we may update $\post{s \mid t} \af{s}$ only as long as the sequence $\{ \post{s \mid t} \af{s} \}_{t = s}^\infty$ exhibits discernible fluctuations, i.e., until, say, $t = s_\varepsilon$, and approximate $\post{s \mid t} \af{s}$ by $\post{s \mid s_\varepsilon} \af{s}$ for all $t \geq s_\varepsilon$. Here $\varepsilon$ is an algorithmic parameter regulating the stopping criterion. 
This idea is explored in the following. 

\subsection{An ideal algorithm}
\label{ssec:alg}

Let $(s, t) \in \nset$ be such that $s \leq t$ and consider the marginal expectation $\post{s \mid t} \af{s} = \E{}[\af{s}(X_s) \mid Y_{0:t} = y_{0:t}]$. By the tower property,  
\begin{equation} \label{eq:smooth:filt}
    \E \left[\af{s}(X_s) \mid Y_{0:t} = y_{0:t} \right] = \E \left[ \tstatfun{s \mid t}(X_t) \mid Y_{0:t} = y_{0:t} \right],
\end{equation}
where
\begin{multline}
    \tstatfun{s \mid t}(x_t) \eqdef \E[ \af{s}(X_s) \mid Y_{0:t-1} = y_{0:t - 1}, X_t = x_t], \\ 
    \quad x_t \in \set{X},  \label{eq:tstat}
\end{multline}
is a statistic appearing frequently in the literature on smoothing; see for instance~\cite{olsson:westerborn:2014b,douc:garivier:moulines:olsson:2010,delmoral:doucet:singh:2009}. Appealingly, reapplying the tower property, the statistics $\{ \tstatfun{s \mid t} \}_{t = s}^\infty$ can be expressed recursively (see~\cite{mongillo:deneve:2008,cappe:2009,delmoral:doucet:singh:2010}) through
\begin{align}
    \tstatfun{s \mid t+1}(x_{t+1}) &= \E[\tstatfun{s \mid t}(X_t) \mid Y_{0:t} = y_{0:t}, X_{t+1}=x_{t+1}] \nonumber \\
    &= \int \tstatfun{s \mid t} (x_t) \, \bk{\post{t}}(x_{t + 1}, \rmd x_t) \nonumber \\
    &= \int \tstatfun{s \mid t} (x_t) \frac{ \hd(x_t, x_{t + 1}) \, \post{t}(\rmd x_t)}{\int \hd(x'_t, x_{t + 1}) \, \post{t}(\rmd x'_t)}, 
    \quad x_{t + 1} \in \set{X}. \label{eq:T:rec}
\end{align}
The recursion is initialised by setting
$$
    \tstatfun{s \mid s}(x_s) \eqdef \af{s}(x_s), \quad x_s \in \set{X}, 
$$
and for completeness we define
$$
    \tstatfun{s \mid u} \equiv 0 \text{ for } u < s.
$$
By \eqref{eq:smooth:filt}, $\post{s \mid t} \af{s} = \post{t} \tstatfun{s \mid t}$, and the target can hence be calculated by applying the filter $\post{t}$ to the function $\tstatfun{s \mid t}$. 

Now the question arises when to stop updating the quantity of interest; indeed, since we are interested in computing the full vector \eqref{eq:tar} but opposed to letting the computational complexity of the algorithm increase with time, we are forced to terminate updating when the fluctuations of the sequence $\{ \post{t} \tstatfun{s \mid t} \}_{t = s}^\infty$ have ceased. 
In the present paper we will stop updating at the time point $s_\varepsilon$ for which the variance of $\tstatfun{s \mid t}$ under the filter $\post{t}$ falls below some given threshold $\varepsilon > 0$ for the first time. After that, we output $\post{s_\varepsilon} \tstatfun{s \mid s_\varepsilon}$ as our estimate of $\post{t} \tstatfun{s \mid t}$ for all $t \geq s_\varepsilon$. This choice can be clearly motivated by \eqref{eq:T:rec}, from which it is clear that once $\tstatfun{s \mid t}$ is close to constant in the support of $\post{t}$, then also $\tstatfun{s \mid t + 1}$ is close to constant everywhere. 

The algorithm can be summarized as follows. 
\begin{itemize}
	\item Initialise by letting $\set{S} \gets \emptyset$. The set $\set{S}$ will keep track of our active estimators. In addition, set the tolerance $\tol$.
	\item For $t \gets 0, 1, 2, 3, \ldots$
	\begin{itemize}
		\item for each $s \in \set{S}$, calculate $\tstatfun{s \mid t}(x_t)$ using~\eqref{eq:T:rec};
		\item let $\set{S} \gets \set{S} \cup \{t\}$, i.e. activate an estimator at time $t$ and set $\tstatfun{t \mid t} \gets \af{t}$;
		\item for each $s \in \set{S}$, calculate the variance $\mathbb{V}_{\post{t}}[\tstatfun{s \mid t}(X_t)]$; if it is smaller than $\tol$, let $\set{S} \gets \set{S} \setminus \{s\}$ and output $\post{t}\tstatfun{s \mid t}$.
	\end{itemize}
\end{itemize}

\subsection{Example: linear Gaussian SSMs}
\label{ssec:kalman}
As mentioned previously, exact computation of the filter and joint-smoothing distributions is possible only for a few specific models. Here we present a Kalman-based version in the case of linear Gaussian SSMs. 

In the linear Gaussian SSMs, an $n_x$-dimensional autoregressive state process is partially observed through $n_y$-dimensional observations.  The model is specified by the equations
\begin{align}
	X_{t+1} &= A X_{t} + U_{t+1}, \\
	Y_t &= B X_t + V_{t},
\end{align}
where $A \in \mathbb{R}^{n_x \times n_x}$ and $B \in \mathbb{R}^{n_y \times n_x}$ and $\{U_t\}_{t \in \nset}$ and $\{V_t\}_{t \in \nset}$ are sequences of mutually independent Gaussian noises with zero mean and covariance matrices $\Sigma_U \in \mathbb{R}^{n_x \times n_x}$ and $\Sigma_V \in \mathbb{R}^{n_y \times n_y}$, respectively. All matrices  are assumed to be pre-specified. Given the sequence $\{ y_t \}_{t \in \nset}$ of observations we wish to estimate \eqref{eq:tar} in the case of \emph{affine} objective functions $\af{s}(x) = \alpha_s^{\intercal} x + \beta_s$, where $\alpha_s \in \mathbb{R}^{n_x}$ and $\beta_s \in \mathbb{R}$ are pre-specified.

In this model each filter $\post{t}$ is Gaussian, and the \emph{Kalman filter} propagates its mean $\mu_t$ and covariance matrix $\Sigma_t$ recursively through time. We calculate the backward kernel \eqref{eq:bk}, which is proportional to the filter at time $s$ times the transition density of the latent Markov chain. Since both these distributions are Gaussian with known mean and covariance matrices, it is easy to show that also the distribution of $X_t$ conditioned on $X_{t+1} = x_{t + 1}$ and $Y_{0:t} = y_{0:t}$ is Gaussian with mean $\mu_{t \mid t+1}(x_{t+1}) = \Sigma_{t \mid t+1} ( A^{\intercal} \Sigma_U^{-1} x_{t+1} + \Sigma_t^{-1} \mu_{t})$ and covariance matrix $\Sigma_{t \mid t+1} = (A^{\intercal} \Sigma_U^{-1} A + \Sigma_t^{-1})^{-1}$. We may hence write down a specific updating procedure for the functions $\{ \tstatfun{s \mid t} \}_{t = s}^\infty$ in this case:
\begin{itemize}
	\item Initialisation: for $t = s$, let $\tstatfun{s \mid s}(x_s) = \af{s}(x_s) = \alpha_s^{\intercal} x_s + \beta_s$.
	\item Proceeding recursively, assume that $\alpha_{s \mid t}$ and $\beta_{s \mid t}$ are of form $\tstatfun{s \mid t}(x_t) = \alpha_{s \mid t}^{\intercal} x_t + \beta_{s \mid t}$; then $\tstatfun{s \mid t+1}(x_{t+1}) = \alpha_{s \mid t+1}^{\intercal} x_{t+1} + \beta_{s \mid t+1}$, where
	\begin{align}
		\alpha_{s \mid t+1}^{\intercal} &= \alpha_{s \mid t}^{\intercal} \Sigma_{t \mid t+1}A^{\intercal} \Sigma_U^{-1},\\
		\beta_{s \mid t+1} &= \alpha_{s \mid t}^{\intercal} \Sigma_{t \mid t+1}A^{\intercal} \Sigma_t^{-1} \mu_t + \beta_{s \mid t}.
	\end{align}
\end{itemize}

The last step of the algorithm---consisting in checking whether the variance of the function above is small enough---is now easily performed by calculating $\mathbb{V}_{\post{t}}[\tstatfun{s \mid t}(X_t)] = \alpha_{s \mid t}^{\intercal} \Sigma_{t} \alpha_{s \mid t}$ and comparing this with some pre-specified threshold $\tol$. This completes all the steps needed for executing the algorithm, and we refer to \autoref{sec:simul} for a numerical illustration. 

Even though the previous method provides an exact implementation of the algorithm, it is limited to a single class of models and specific target functions. To move beyond this simplified setting we need to rely on approximations, and this will be discussed in the next section.

\section{Particle-based online marginal smoothing}
\label{sec:pf}

As mentioned previously, exact computation of the filter distributions---and hence the backward kernels---is possible only in a few specific cases. In the general case we will approximate these distributions using particle filters, which are recalled in the following. 

\subsection{Particle filters}

A particle filter propagates recursively a set of particles with associated weights in order to approximate the filter distribution flow $\{\post{t}\}_{t \in \nset}$ given the sequence $\{y_t\}_{t \in \nset}$. 

We describe recursively the most basic particle filter---the bootstrap filter~\cite{gordon:salmond:smith:1993}---and assume that we have at hand a sample $\{(\epart{t}{i},\wgt{t}{i})\}_{i = 1}^{\N}$ of particles (the $\epart{t}{i}$) and associated weights (the $\wgt{t}{i}$) targeting the filter distribution $\post{t}$ in the sense that for all $\post{t}$-integrable functions $f$, 
\begin{equation}
	\sum_{i=1}^{\N}\frac{\wgt{t}{i}}{\wgtsum{t}}\testf(\epart{t}{i}) \backsimeq \post{t}\testf, 
\end{equation}
where $\wgtsum{t} \eqdef \sum_{i=1}^{\N}\wgt{t}{i}$ denotes the total weight and ``$\backsimeq$'' means that the estimator on the left hand side is \emph{consistent}, i.e., converges in probability to the right hand side, as $\N \to \infty$. To form a new weighted sample $\{(\wgt{t+1}{i}, \epart{t+1}{i})\}_{i=1}^{\N}$ targeting the subsequent filter distribution $\post{t+1}$, a two-step procedure is applied. First, the particles are resampled by drawing randomly a set of $\N$ independent indices $\{I_{t+1}^{i}\}_{i=1}^{\N}$ from the categorical distribution induced by the probabilities proportional to the weights $\{\wgt{t}{i}\}_{i=1}^{\N}$, an operation denoted by $I_{t+1}^{i} \sim \probdist( \{\wgt{t}{i}\}_{i=1}^{\N})$, $i \in \{1, \ldots, \N\}$.

Second, the resampled particles are moved conditionally independently according to the dynamics of the state process, i.e., for all $i \in \{1, \ldots, \N \}$, 
\begin{align}
    \epart{t+1}{i} \sim \hk{}(\epart{t}{I_{t+1}^i}, \cdot). 
\end{align}
Finally, new importance weights are computed according to
\begin{align}
     \wgt{t+1}{i} = \md{t+1}(\epart{t+1}{i})
\end{align}
for all $i \in \{1, \ldots, \N \}$.
 
Initialisation is carried through by drawing $\epart{0}{i} \sim \Xinit{}$ and setting $\wgt{0}{i} = \md{0}(\epart{0}{i})$ for all $i \in \{1, \ldots, \N\}$.

We summarise the procedure in~\autoref{alg:bpf}, and denote by ``$\{(\epart{t+1}{i},\wgt{t+1}{i}) \}_{i=1}^{\N} \gets \PF(\{(\epart{t}{i},\wgt{t}{i})\}_{i=1}^{\N}, y_{t + 1})$'' one application of~\autoref{alg:bpf}. (By convention we let $\{ (\epart{0}{i},\wgt{0}{i}) \}_{i = 1}^\N \gets \PF(\{(\epart{-1}{i},\wgt{-1}{i})\}_{i=1}^{\N}, y_0)$ denote the initial step.)

\begin{algorithm}[htb]
    \caption{Bootstrap particle filter}
    \label{alg:bpf}
    \begin{algorithmic}[1]
    	\Require A weighted sample $\{(\epart{t}{i},\wgt{t}{i})\}_{i=1}^{\N}$ targeting the filter distribution $\post{t}$
    	\For{$i = 1 \to \N$}
    		\State draw $I_{t + 1}^{i} \sim \probdist(\{\wgt{t}{\ell}\}_{\ell = 1}^{\N})$;
    		\State draw $\epart{t+1}{i} \sim \hk(\epart{t}{I_{t + 1}^i},\cdot)$;
    		\State set $\wgt{t+1}{i} = \md{t + 1}(\epart{t+1}{i})$;
    	\EndFor
    	\State \Return $\{ (\epart{t+1}{i},\wgt{t+1}{i}) \}_{i=1}^{\N}$
    \end{algorithmic}
\end{algorithm}

%
%

So far we have only considered estimation of the filter distributions. The rest of this section will be devoted to particle approximation of the marginal smoothing distributions. 

For all $i \in \{1, \ldots, \N \}$, define recursively genealogical indices $\{G_{s \mid t}^i\}_{s = 0}^{t}$ by $G_{t \mid t}^i = i$ and $G_{s - 1 \mid t}^{i} = I_s^{G_{s \mid t}^{i}}$. The set $\{ (\epart{0}{G_{0 \mid t}^i}, \ldots, \epart{t}{G_{t \mid t}^i}) \}_{i = 1}^\N$ is often referred to as the \emph{genealogical tree} of the particles, and it is easy to show that the genealogical tree may, together with the importance weights $\{\wgt{t}{i}\}_{i=1}^{\N}$, be used for estimating the joint-smoothing distribution. In particular, 
\begin{align}
      \sum_{i=1}^{\N} \frac{\wgt{t}{i}}{\wgtsum{t}} \af{s}(\epart{s}{G_{s \mid t}^{i}}) \backsimeq \post{s \mid t}\af{s},   
      \label{eq:poor:man:smoother}
\end{align}
which means a given marginal smoothing distribution $\post{s \mid t}$ can be approximated using the weighted sample formed by the \emph{time} $s$ \emph{ancestors} of $\{ \epart{t}{i} \}_{i = 1}^N$. We refer to this estimator as the \emph{Poor man's smoother}. A well-known problem with the Poor man's smoother is that the repeated resampling operations of the particle filter always deplete the genealogical tree in the long run; thus, sooner or later, for some $s < t$ and some $i_0$, $G_{u \mid t}^{i_0} = G_{u \mid t}^{j}$ for all $j \in \{1, \ldots, \N\}$ and all $u \in \{0, \ldots, s\}$, implying that the estimates of $\post{u \mid t} \af{u}$, $u \in \{0,\ldots,s\}$, will be based on only a single particle path. The work~\cite{jacob:murray:rubenthaler:2013} establishes, under assumptions requiring typically the state space to be a compact set, a bound on the expected distance from the last generation to the most recent common ancestor that is proportional to $\N \log(\N)$ and \emph{uniform} in time. Thus, the number of active, unique particles in the estimator~\eqref{eq:poor:man:smoother} tends to one as $t$ increases, leading to a depleted and impractical estimator.

\subsection{Fixed-lag smoothing}\label{ssec:fls}

To remedy the problem of particle lineage degeneracy, a fixed-lag smoother~\cite{kitagawa:sato:2001} can be used. The idea is to approximate the marginal smoothing distribution $\post{s \mid t}$ by the distribution $\post{s \mid \lag{t}}$, where $\lag{t} = (s + \Delta) \wedge t$ for some pre-specified lag $\Delta \in \nsetpos$. In this case, using the notation above, we get the biased approximation
$$
	\post{s \mid t} \af{s} \approx \sum_{i = 1}^{\N} \frac{\wgt{\lag{t}}{i}}{\wgtsum{\lag{t}}}\af{s}\big( \epart{s}{G_{s \mid \lag{t}}^i} \big),
$$
where $G_{s \mid \lag{t}}^{i}$ is defined as above.

The approach requires suitable design of the lag $\Delta$, and we face here a classical bias-variance tradeoff: if $\Delta$ is too small, then the forgetting of the model has not kicked in, and the discrepancy between the distributions is going to be large (leading to high bias); on the other hand, if $\Delta$ is too large, then, by path degeneracy, the estimate will be depleted (leading to high variance). The optimal choice of lag depends on the mixing of the SSM, and~\cite{olsson:cappe:douc:moulines:2006} proposes an optimal choice of $\Delta$ as $\lceil c \log(t) \rceil$, where the constant $c$ depends on the mixing. This is problematic since it is hard to calculate, and even estimate, the mixing of an SSM. Thus, designing properly the lag is indeed a non-trivial task. The lag-based particle estimator that we propose in the next section relies again on forgetting-based arguments, but adapts the lag in a completely automatic manner.  

\subsection{The adaptive-lag smoother}

We present here a particle-based version of the ideal algorithm in~\autoref{ssec:alg}, which can be thought of as an adaptive-lag smoother. In this algorithm we employ novel techniques for updating particle estimates of the functions $\tstatfun{s \mid t}$, and the adaptive lag can be thought of as the number of steps that each function is updated before the stopping criterion triggers truncation. The truncation depends heavily on the mixing of the model, but is now determined in an adaptive manner. 

The critical step in the algorithm presented in~\autoref{ssec:alg} is the need of estimating and updating the statistics $\tstatfun{s \mid t}$ through the recursion~\eqref{eq:T:rec}. We proceed by induction and assume that we have at hand a set $\{ \tstattil{i}{s \mid t} \}_{i=1}^{\N}$ of estimates of $\{ \tstatfun{s \mid t}(\epart{t}{i}) \}_{i=1}^{\N}$. Proceeding as in \cite{delmoral:doucet:singh:2009}, these estimates are updated to estimates $\{ \tstattil{i}{s \mid t+1} \}_{i=1}^{\N}$ of $\{ \tstatfun{s \mid t+1}(\epart{t+1}{i}) \}_{i=1}^{\N}$ by replacing, in \eqref{eq:T:rec}, $\post{t}$ by a particle approximation, yielding 
\begin{align}
    \tstattil{i}{s \mid t+1} = \sum_{j=1}^{\N} \frac{\wgt{t}{j}\hd({\epart{t}{j},\epart{t+1}{i}})}{\sum_{\ell = 1}^{\N} \wgt{t}{\ell} \hd(\epart{t}{\ell},\epart{t+1}{i})}\tstattil{j}{s \mid t}, \  i \in \{1, \ldots, \N \}, \label{eq:FFBSm:update}
\end{align}
where the ratio is a particle approximation of the backward kernel~\eqref{eq:bk}, and the particle approximation $\sum_{i = 1}^\N \wgt{t}{i} \tstattil{i}{s \mid t} / \wgtsum{t} \backsimeq \post{s \mid t} \af{s}$. Casting the recursion \eqref{eq:FFBSm:update} into the ideal algorithm in \autoref{sec:algo} yields the following procedure: 

\begin{itemize}
    \item Initialise by letting $\set{S} \gets \emptyset$ and setting the tolerance $\tol$.
    \item For $t \gets 0, 1, 2, 3,\ldots$
    \begin{itemize}
        \item run $\{(\epart{t}{i},\wgt{t}{i})\}_{i=1}^{\N} \gets \PF{}( \{(\epart{t-1}{i},\wgt{t-1}{i})\}_{i=1}^{\N}, y_t)$;
        \item for each $s \in \set{S}$ and $i \in \{1, \ldots, \N\}$, calculate $\tstattil{i}{s \mid t}$ using~\eqref{eq:FFBSm:update};
        \item let $\set{S} \gets \set{S} \cup \{t\}$ and $\tstattil{i}{t \mid t} \gets \af{t}(\epart{t}{i})$ for all $i \in \{1, \ldots, \N\}$.
        \item for each $s \in \set{S}$, if
        $$
         \sum_{i=1}^{\N} \frac{\wgt{t}{i}}{\wgtsum{t}} \left( \tstattil{i}{s \mid t} -  \sum_{j=1}^{\N} \frac{\wgt{t}{j}}{\wgtsum{t}} \tstattil{j}{s \mid t} \right)^2 < \tol,
        $$
        then let $\set{S} \gets \set{S} \setminus \{s\}$ and output 
        $\sum_{i=1}^{\N} \wgt{t}{i} \tstattil{i}{s \mid t} / \wgtsum{t}$ as an estimate of $\post{s \mid t'} h_s$ for all $t' \geq t$.
    \end{itemize}
\end{itemize}

A drawback with the updating formula \eqref{eq:FFBSm:update} is that it requires a sum of $\N$ terms to be computed for each particle, which yields an overall $\ordo{}(\N^2)$ computational complexity. Needless to say, this is impractical when $\N$ is large. 

To reduce the computational burden, we proceed as in the PaRIS \cite[Alg.~2]{olsson:westerborn:2014b} and replace the right hand side of~\eqref{eq:FFBSm:update}, which can be interpreted as an expectation, by a Monte Carlo estimate. More precisely, assuming that we have at hand a set $\{\tstat[i]{s \mid t}\}_{i=1}^{\N}$ of estimates of $\{\tstatfun{s \mid t}(\epart{t}{i})\}_{i=1}^{\N}$, we replace \eqref{eq:FFBSm:update} by the mean
\begin{equation} \label{eq:PaRIS-type:update}
    \tstat[i]{s \mid t+1} = \frac{1}{\K} \sum_{j=1}^{\K} \tstat[\bi{t}{i}{j}]{s \mid t},
\end{equation}
where $\{\bi{t}{i}{j} \}_{j = 1}^{\K}$ are conditionally independent draws from $\probdist{}(\{ \wgt{t}{\ell}\hd(\epart{t}{\ell}, \epart{t+1}{i}) \}_{\ell=1}^{\N})$ and $\K$ is a precision parameter. Such draws can most often be produced at low computational cost using rejection sampling. Indeed, assume that the transition density $\hd{}$ is uniformly bounded by some constant $\hkup$, i.e., $\hd{}(x, x') \leq \hkup$ for all $(x, x') \in \set{X}^2$; then, following \cite{douc:garivier:moulines:olsson:2010}, a draw $J$ from $\probdist{}(\{ \wgt{t}{\ell}\hd(\epart{t}{\ell}, \epart{t+1}{i}) \}_{\ell=1}^{\N})$ can be produced by repeating the following steps until acceptance:
\begin{itemize}
    \item[(1)] draw $J \sim \probdist{}(\{\wgt{t}{i}\}_{i=1}^{\N})$;
    \item[(2)] accept $J$ with probability $\hd{}(\epart{t}{J},\epart{t + 1}{i})/\hkup{}$.
\end{itemize}
Interestingly, under the mixing assumptions given in \autoref{sec:theory} it is possible to show that the expected number of trials required for sampling all the indices $\{\bi{t}{i}{j}\}_{j=1}^{\K}$ is linear in $\K$; see~\cite[Thm. 10]{olsson:westerborn:2014b}. This yields an $\ordo(\N \K)$ algorithm, and as we will see below, $\K$ can be kept at a very low value (say, $\K = 2$).   

The variance of $\tstatfun{s \mid t}$ under $\post{t}$ is estimated using 
$$
    \varlim[part]{s \mid t} \eqdef \sum_{i=1}^{\N}\frac{\wgt{t}{i}}{\wgtsum{t}}\left( \tstat[i]{s \mid t} - \sum_{\ell=1}^{\N} \frac{\wgt{t}{\ell}}{\wgtsum{t}} \tstat[\ell]{s \mid t} \right)^2,
$$
and the updating procedure is stopped if $\varlim[part]{s \mid t} < \tol$, where $\tol$ is some pre-chosen tolerance. Letting 
\begin{align}
     s_{\tol}^{\N}(t) \eqdef \min \{u \geq s : \varlim[part]{s \mid u} < \tol \} \wedge t,  
     \label{eq:part:stop}
 \end{align}
 we return the estimator 
\begin{align}
    \postafl[part]{s \mid t} \af{s} \eqdef \sum_{i=1}^{\N} \frac{\wgt{s_{\tol}^{\N}(t)}{i}}{\wgtsum{s_{\tol}^{\N}(t)}} \tstat[i]{s \mid s_{\tol}^{\N}(t)}
\end{align}
of $\post{s \mid t} \af{s}$. The algorithm is presented in detail in~\autoref{alg:afls}. 

\begin{algorithm}[htb]
    \caption{Adaptive-lag smoother}
    \label{alg:afls}
    \begin{algorithmic}[1]
        \State set $\set{S} \gets \{0\}$;
        \State run $\{ (\epart{0}{i}, \wgt{0}{i})\}_{i=1}^{\N} \gets \PF{}( \{(\epart{-1}{i}, \wgt{-1}{i})\}_{i=1}^{\N}, y_0)$;
        \For{$i = 1 \to \N$}
            \State set $\tstat[i]{0 \mid 0} \gets \addf{0}(\epart{0}{i})$;
        \EndFor
        \For{$t \gets 1, 2, 3, \ldots$}
            \State run $\{ (\epart{t}{i}, \wgt{t}{i}) \}_{i = 1}^\N \gets \PF( \{ (\epart{t-1}{i}, \wgt{t-1}{i}) \}_{i = 1}^\N, y_t)$;
            \For{$i = 1 \to \N$}
                \For{$j = 1 \to \K$}
                    \State draw $\bi{t}{i}{j} \sim \probdist( \{ \wgt{t-1}{\ell} q(\epart{t-1}{\ell}, \epart{t}{i} ) \}_{\ell = 1}^\N )$;
                \EndFor
                \For{$s \in \set{S}$}
                    \State set $\tstat[i]{s\mid t} \gets \K^{-1} \sum_{\k = 1}^{\K} \tstat[\bi{t}{i}{\k}]{s \mid t-1}$;
                \EndFor

                \State set $\tstat[i]{t \mid t} \gets \af{t}(\epart{t}{i})$;
            \EndFor

            \State set $\set{S} \gets \set{S} \cup \{t\}$;
            \For{$s \in \set{S}$}
                \If{$\varlim[part]{s \mid t} < \tol$}
                    \State set $\postafl[part]{s \mid t'} h_s \gets \sum_{i=1}^{\N} \wgt{t}{i} \tstat[i]{s \mid t} / \wgtsum{t}$ for all $t' \geq t$;
                    \State set $\set{S} \gets \set{S} \setminus \{s\}$.
                \EndIf
            \EndFor
        \EndFor
    \end{algorithmic}
\end{algorithm}

\begin{remark}
In \autoref{alg:afls} we let, on Line~7, the underlying particles be propagated by means of the standard bootstrap particle filter for simplicity. However, from a methodological point of view, running the proposed algorithm requires only access to a  sequence of consistent samples---with or without weights---targeting the filter distribution flow, and it is, in practice, of subordinate importance how these samples are produced. This is relevant for, e.g., high-dimensional scenarios or scenarios with highly informative observations, where alternative particle filters, such as \emph{feedback particle filters} \cite{yang:mehta:meyn:2013} or \emph{auxiliary particle filters} \cite{pitt:shephard:1999}, may improve estimation significantly.  
\end{remark}

\subsection{Designing algorithmic parameters}
\begin{figure*}
\begin{minipage}[t]{0.45\linewidth}
\centering
\centerline{\begin {tikzpicture}[-latex ,auto ,node distance = 0.3 cm,
semithick ,
state/.style ={ circle ,
fill = black,black , text=black , minimum width =0.5 cm, radius = .7}]
\tikzstyle{eCirc}=[circle, minimum width = 0.5cm, radius = .7, draw = black, pattern=horizontal lines, pattern color=black]
\node[state, fill = lightgray] (A1){};
\node[eCirc] (A2) [above =of A1]{};
\node[state, fill = lightgray] (A3) [above =of A2]{};
\node[state, fill = lightgray] (B1) [right =of A1]{};
\node[state] (B2) [right =of A2]{};
\node[state, fill = lightgray] (B3) [right =of A3]{};
\node[state] (C1) [right =of B1]{};
\node[state, fill = lightgray] (C2) [right =of B2]{};
\node[state, fill = lightgray] (C3) [right =of B3]{};
\node[state, fill = lightgray] (D1) [right =of C1]{};
\node[state] (D2) [right =of C2]{};
\node[state] (D3) [right =of C3]{};
\node[state] (E1) [right =of D1]{};
\node[state] (E2) [right =of D2]{};
\node[state] (E3) [right =of D3]{};
\path[draw,dashed,color = lightgray] (B1) -- (A1);
\path[draw,dashed,color = black] (B2) -- (A2);
\path[draw,dashed,color = lightgray] (B3) -- (A2);
\path[draw,dashed,color = black] (C1) -- (B2);
\path[draw,dashed,color = lightgray] (C2) -- (B3);
\path[draw,dashed,color = lightgray] (C3) -- (B2);
\path[draw,dashed,color = lightgray] (D1) -- (C2);
\path[draw,dashed,color = black] (D2) -- (C1);
\path[draw,dashed,color = black] (D3) -- (C1);
\path[draw,dashed,color = black] (E1) -- (D2);
\path[draw,dashed,color = black] (E2) -- (D2);
\path[draw,dashed,color = black] (E3) -- (D3);
\end{tikzpicture}}
\centerline{\scriptsize (a) $\K = 1$}
\end{minipage}
\hfill
\begin{minipage}[t]{0.45\linewidth}
\centering
\centerline{\begin {tikzpicture}[-latex ,auto ,node distance = 0.4 cm,
semithick ,
state/.style ={ circle ,
fill = black,black , text=black , minimum width =0.5 cm, radius = .7}]
\tikzstyle{eCirc}=[circle, minimum width = 0.5cm, radius = .7, draw = black, pattern=horizontal lines, pattern color=black]
\node[eCirc] (A1){};
\node[eCirc] (A2) [above =of A1]{};
\node[eCirc] (A3) [above =of A2]{};
\node[state] (B1) [right =of A1]{};
\node[state] (B2) [right =of A2]{};
\node[state] (B3) [right =of A3]{};
\node[state] (C1) [right =of B1]{};
\node[state] (C2) [right =of B2]{};
\node[state, fill = lightgray] (C3) [right =of B3]{};
\node[state, fill = lightgray] (D1) [right =of C1]{};
\node[state] (D2) [right =of C2]{};
\node[state] (D3) [right =of C3]{};
\node[state] (E1) [right =of D1]{};
\node[state] (E2) [right =of D2]{};
\node[state] (E3) [right =of D3]{};
\path[draw,dashed,color = black] (B1) -- (A1);
\path[draw,dashed,color = black] (B2) -- (A2);
\path[draw,dashed,color = black] (B3) -- (A2);
\path[draw,dashed,color = black] (C1) -- (B2);
\path[draw,dashed,color = black] (C2) edge [bend right] (B3);
\path[draw,dashed,color = lightgray] (C3) -- (B2);
\path[draw,dashed,color = lightgray] (D1) -- (C2);
\path[draw,dashed,color = black] (D2) -- (C1);
\path[draw,dashed,color = black] (D3) -- (C1);
\path[draw,dashed,color = black] (E1) -- (D2);
\path[draw,dashed,color = black] (E2) -- (D2);
\path[draw,dashed,color = black] (E3) -- (D3);
\path[draw,dashed,color = black] (B1) -- (A3);
\path[draw,dashed,color = black] (B2) -- (A1);
\path[draw,dashed,color = black] (B3) -- (A3);
\path[draw,dashed,color = black] (C1) -- (B1);
\path[draw,dashed,color = black] (C2) edge [bend left] (B3);
\path[draw,dashed,color = lightgray] (C3) -- (B1);
\path[draw,dashed,color = lightgray] (D1) -- (C1);
\path[draw,dashed,color = black] (D2) -- (C2);
\path[draw,dashed,color = black] (D3) -- (C2);
\path[draw,dashed,color = black] (E1) -- (D3);
\path[draw,dashed,color = black] (E2) -- (D3);
\path[draw,dashed,color = black] (E3) -- (D2);
\end{tikzpicture}}
\centerline{\scriptsize (b) $\K = 2$}
\end{minipage}
\caption{Genealogical traces corresponding to backward simulation in the adaptive-lag smoothing algorithm. Columns of nodes refer to different particle populations (with $\N = 3$) at different time points (with time increasing rightward) and arrows indicate connections through the backward draws in the algorithm. The striped particles are in the final estimator for the marginal smoothing distribution when $s=0$ and $t=4$, black-colored particles are in the support of the historical trajectories of the final estimator, while gray-colored ones are inactive.}
\label{fig:IllDeg}
\end{figure*}
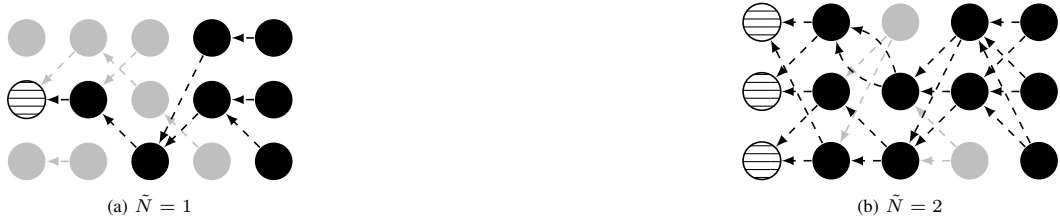

In \autoref{alg:afls}, the parameters $\tol$ and $\K$ are set by the user. Interestingly, \cite{olsson:westerborn:2014b} establishes that the PaRIS is (i) consistent for all fixed $\K \in \nsetpos$ and (ii) numerically stable only if $\K \geq 2$. The latter is illustrated by~\autoref{fig:IllDeg}, from which it is clear that using $\K = 1$ leads to a degeneracy phenomenon that is reminiscent of the degeneracy of the genealogical tree. As it is clear from the same figure, this phenomenon is avoided in the case $\K = 2$. This distinction between the cases $\K = 1$ and $\K \geq 2$ is also present in the central limit theorem in~\cite[Thm. 8]{olsson:westerborn:2014b}, where, in the marginal smoothing case, a time uniform $\mathcal{O}(1 + 1/\K)$ bound on the asymptotic variance is obtainable only in the case $\K \geq 2$. As suggested by this bound, there is no gain in using a too large precision $\K$, and typically $\K = 2$ provides a satisfactory accuracy in simulations. 

When it concerns the tolerance $\tol$, using a smaller tolerance implies a larger lag, implying in turn more accurate estimates (and conversely). However, a larger lag requires a larger bank of active estimators, increasing in turn the computational time and memory requirement. This trade-off is studied in more detail in~\autoref{sec:theory} and \autoref{sec:simul}.


\section{Theory}
\label{sec:theory}

\subsection{Convergence of the sample variance criterion}
We start off the theoretical analysis by studying the asymptotics (as $\N \rightarrow \infty$) of the sample variances $\varlim[part]{s \mid t}$. 
 
The analysis will be carried through under the following assumptions. 
\begin{assumption} \label{ass:bnd}
For all $t \in \nset$, $\supn{\md{t}} < \infty$. Moreover, there exists a constant $\hbdnt < \infty $ such that $\oscn{\af{t}} < \hbdnt$ for all $t \in \nset$.
\end{assumption}

Note that the first part of \autoref{ass:bnd} implies finiteness of the particle weights. The following auxiliary result establishes the convergence of each sample variance criterion to a deterministic limit. 

\begin{lemma}\label{lem:var-conv}
	Let \autoref{ass:bnd} hold. Then for all $(s, t) \in \nset^2$ such that $s \leq t$, it holds, as $\N \to \infty$, 
	\begin{equation}
		\sum_{i=1}^{\N} \frac{\wgt{t}{i}}{\wgtsum{t}}\left( \tstat[i]{s \mid t} - \sum_{\ell =1}^{\N} \frac{\wgt{t}{\ell}}{\wgtsum{t}}\tstat[\ell]{s \mid t}\right)^2 \convp \varlim{s \mid t}, 
	\end{equation}
	where
	\begin{align}
		\varlim{s \mid t} \eqdef \post{t}\{ (\tstatfun{s \mid t} - \post{t} \tstatfun{s \mid t})^2 \} + \sqc{s, t}
	\end{align}
	with $\sqc{s,t}$ being defined in \autoref{sec:append}, Eqn.~\eqref{eq:def:eta}.  
\end{lemma}

Second, we bound the limiting variance criterion. This calls for the following \emph{strong mixing assumptions}.
\begin{assumption} \label{ass:strong:mixing}
\ 
\begin{itemize}
\item[(i)] There exist $0 < \hklow{} < \hkup{} < \infty$ such that $\hklow{} < \hd(x,x') < \hkup{}$ for all $(x,x') \in \set{X}^2$. 
\item[(ii)] There exist $0 < \mdlow{} < \mdup{} < \infty$ such that for all $t \in \nset$, $\mdlow < \md{t}(x) < \mdup$ for all $x \in \set{X}$.
\end{itemize}
\end{assumption}

Under \autoref{ass:strong:mixing}(i), define $\mr \eqdef 1 - \hklow / \hkup$.

Assumptions similar to \autoref{ass:strong:mixing} appear frequently in the literature (see, e.g., \cite{delmoral:2004}) and require typically the state space of the hidden chain to be a compact set. 

\begin{theorem}\label{cor:bounded}
	Under Assumptions~1 and 2, it holds, for all $(s, t) \in \nset^2$ such that $s \leq t$ and all $\K \in \nsetpos$, 
	\begin{equation}
		\varlim{s \mid t} \leq \hbdnt^2  
		 \begin{cases}
			c_1 \mr^{2(t-s)} + c_2 \K^{-(t-s)} & \text{if } \K \mr^{2} \neq 1, \\
			\mr^{2(t-s)} + c_3 (t-s) \K^{-(t-s)} & \text{if } \K \mr^{2} = 1,
		\end{cases} \label{eq:var_bound}
	\end{equation}
	where the constants $c_1, c_2$ and $c_3$ are independent of $s$ and $t$.
\end{theorem}

The first term in the bound \eqref{eq:var_bound} is related to the mixing of the SSM, and since $\mr \in (0, 1)$ this term tends to zero geometrically fast as $t$ grows. The second term is related to the Monte Carlo error induced by the PaRISian update. Here we clearly see that it is required that $\K \geq 2$ in order for this term to vanish as $t$ increases. In that case, $\varlim{s \mid t} \to 0$ as $t \to \infty$.

\subsection{Convergence of the estimator}

In order to derive the asymptotic limit of our estimator we introduce the following notation. Let $s_{\varepsilon}(t) \eqdef \min\{u \geq s : \varlim{s \mid u} < \tol\} \wedge t$, which can be understood as the limit of $s_{\tol}^{\N}(t)$ (defined in \eqref{eq:part:stop}). In addition, let $\postafl{s \mid t} \eqdef \post{s \mid s_{\varepsilon}(t)}$ and notice that this measure differs slightly from the adaptive-lag approximation delivered by the ideal algorithm in \autoref{ssec:alg}, since the limiting variance $\varlim{s \mid t}$ is not equal to $\mathbb{V}_{\post{t}}[\tstatfun{s \mid t}(X_t)]$; recall that the former also has an additional term $\sqc{s, t}$ corresponding to the Monte Carlo approximation of the backward kernel.

As expected and as established by the following theorem, $\postafl{s \mid t}$ is indeed the asymptotic limit of the proposed estimator. 

\begin{theorem}\label{thm:convp}
	Let Assumptions 1 and 2 hold. Then for all $(s, t) \in \nset^2$ such that $s \leq t$ and all bounded measurable functions $\af{s}$, as $\N \to \infty$, 
	\begin{align}
		\postafl[part]{s \mid t}\af{s} \convp \postafl{s \mid t}\af{s}.
	\end{align}
\end{theorem}




\subsection{Bound on asymptotic memory requirement}

Finally, we show that the memory requirement of the algorithm stays, in the asymptotic regime, uniformly bounded in $t$, which was a requirement in the problem statement. Asymptotically, the estimate of $\post{s \mid t} \af{s}$ is still under construction at time $t$ if $\varlim{s \mid t} \geq \tol$ for all $t \in \{s,\ldots,t\}$. Thus, let   
$$
\act{t} \eqdef \sum_{s=0}^{t} \prod_{u = s}^{t} \1{\{\varlim{s \mid u} \geq \tol\}}
$$
be the number of active adaptive-lag estimators at time $t$ in the asymptotic regime. 
\begin{theorem}\label{thm:activeEstimates}
	Under \autoref{ass:strong:mixing}, for all $\K \geq 2$, 
	$$
        		\act{t} \leq \tfrac{\log(\tol / \{\hbdnt^2 d(\mr, \K) \})}{\log (\mr^{2} \vee \K^{-1})}, 
	$$
	where $d(\mr, \K) > 0$ depends on $\mr$ and $\K$ only.
\end{theorem}

We remark that the bound in \autoref{thm:activeEstimates} is uniform in time, implying a uniformly bounded memory requirement of the algorithm, at least in the asymptotic regime. Moreover, we note that the bound in \autoref{thm:activeEstimates} has an $\mathcal{O}(- \log \tol)$ term. 


\section{Simulations}
\label{sec:simul}
We benchmark the algorithm on two different models. First we consider a linear Gaussian SSM, which enables computation of the exact distributions using the disturbance smoother Kalman smoother (see, e.g., \cite[sec. 5.2.4]{cappe:moulines:ryden:2005}). The second model is the now classical stochastic volatility model proposed in~\cite{hull:white:1987}.

\subsection{Linear Gaussian SSM} 
\label{sub:linear_gaussian_state_space_model}
Consider a linear Gaussian SSM given by the following set of equations:
\begin{align}
	X_{t+1} &= a X_t + \sigma_U U_{t+1}, \\
	Y_t &= b X_t + \sigma_V V_t, 
\end{align}
where $\{U_t\}_{t \geq 0}$ and $\{V_t\}_{t \geq 0}$ are independent sequences of mutually independent standard Gaussian noise variables. Initially, $X_0 \sim \mathcal{N}(0, \sigma_V^2/(1 - a^2))$. We consider smoothed means, corresponding to the objective functions $\af{s} = \operatorname{id}$ for all $s \in \nset$. We simulate a data record comprising $201$ observations  $y_{0:200}$ from the the model parameterised by $(a, b, \sigma_U, \sigma_V) = (.95, .5, .5 ,2)$. Using tolerances $\tol \in \{.5, .2, .1, 10^{-3} \}$ and $(\N,\K) = (400, 2)$ we performed $100$ independent runs of the algorithm with the same input data. In addition, the Kalman version of the ideal algorithm (as presented in~\autoref{ssec:kalman}) was run with the same tolerances as the particle version. 

\begin{figure}
\begin{minipage}{0.95\linewidth}
	\includegraphics[width = \linewidth]{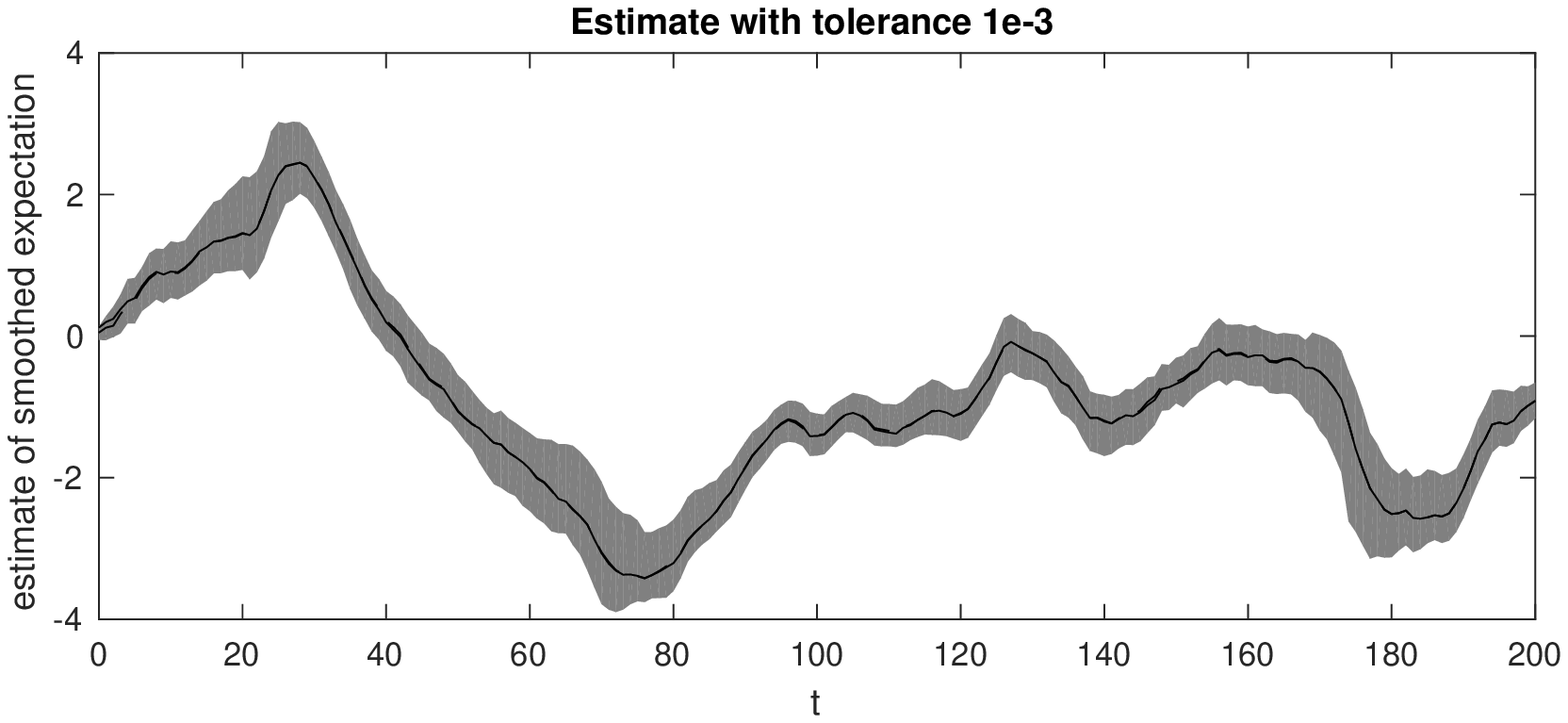}
\end{minipage}
	\begin{minipage}{0.95\linewidth}
	\includegraphics[width = \linewidth]{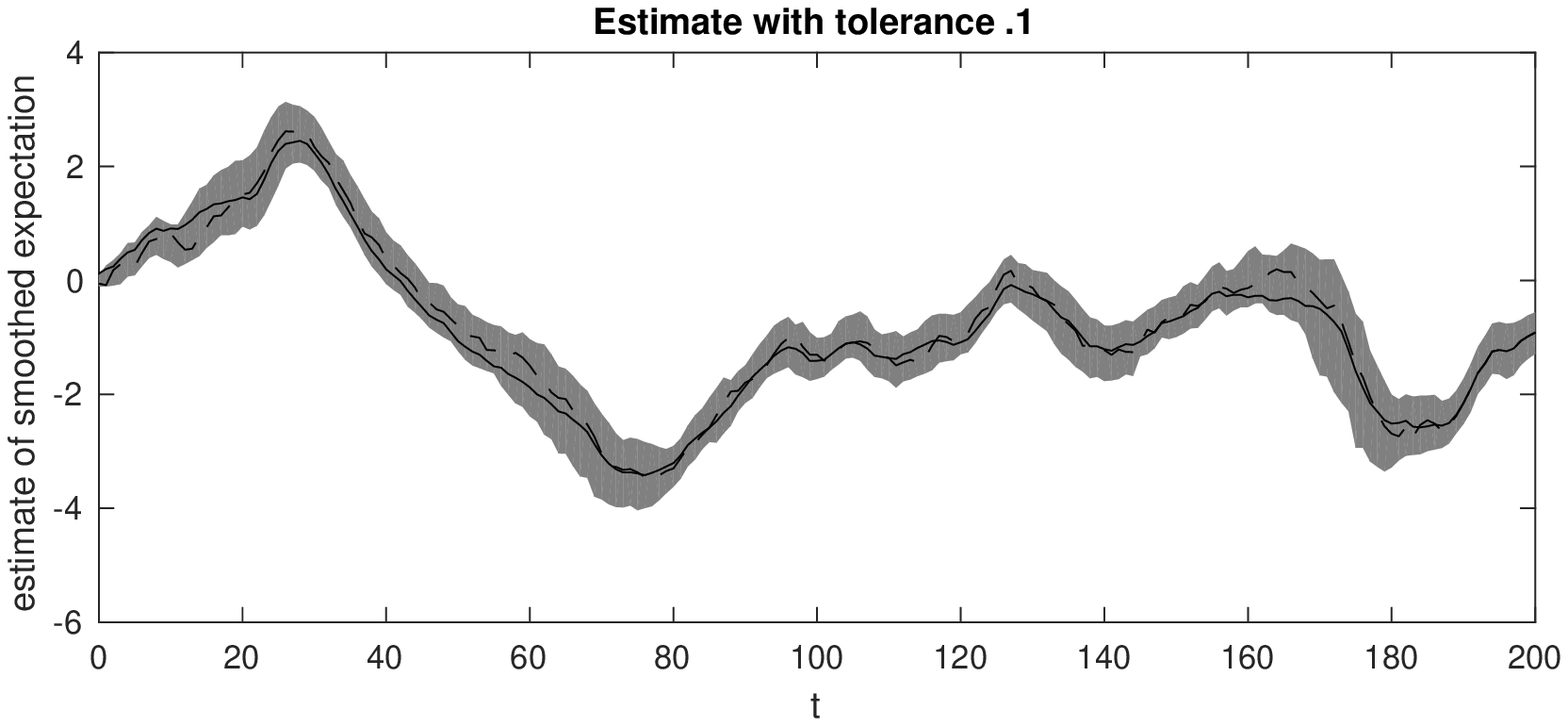}
\end{minipage}
\begin{minipage}{0.95\linewidth}
	\includegraphics[width = \linewidth]{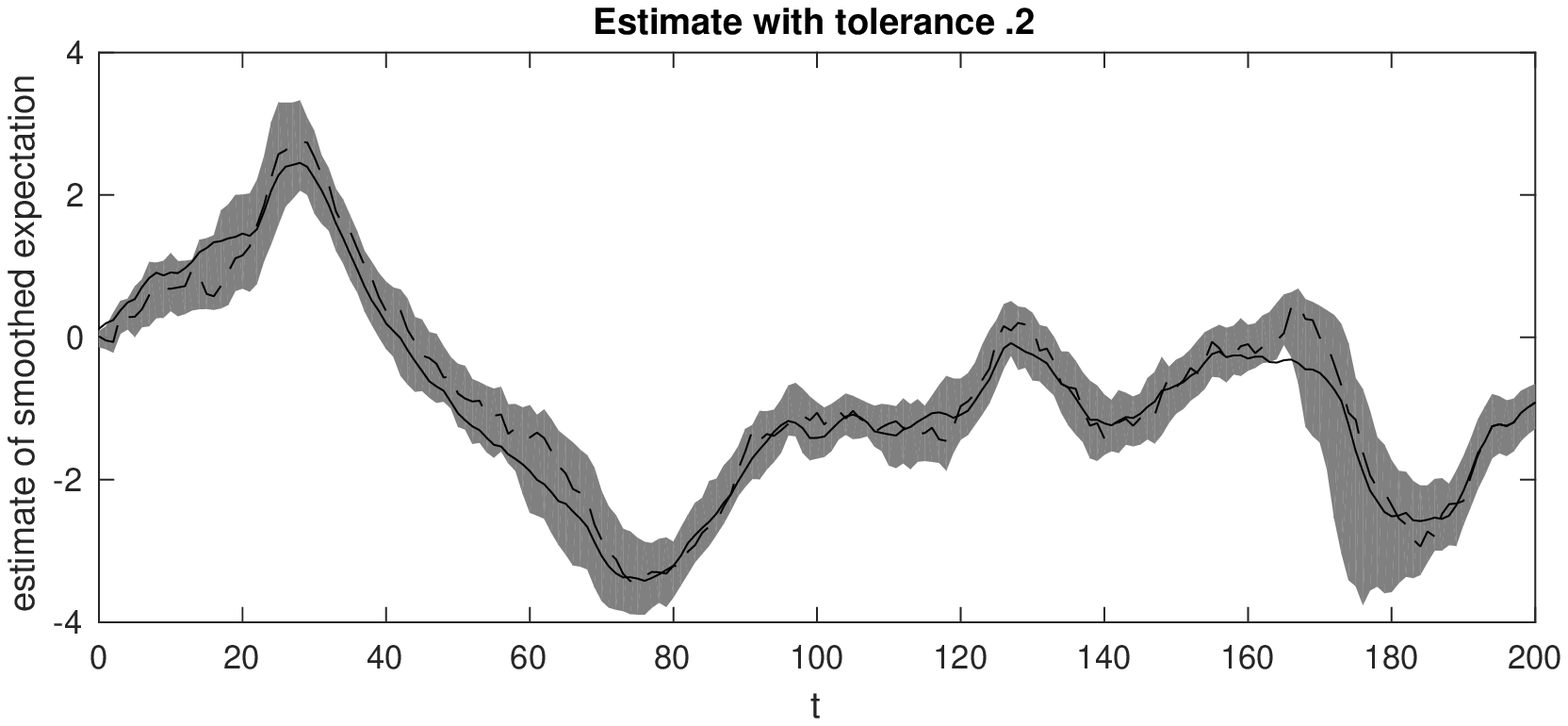}
\end{minipage}
\begin{minipage}{0.95\linewidth}
	\includegraphics[width = \linewidth]{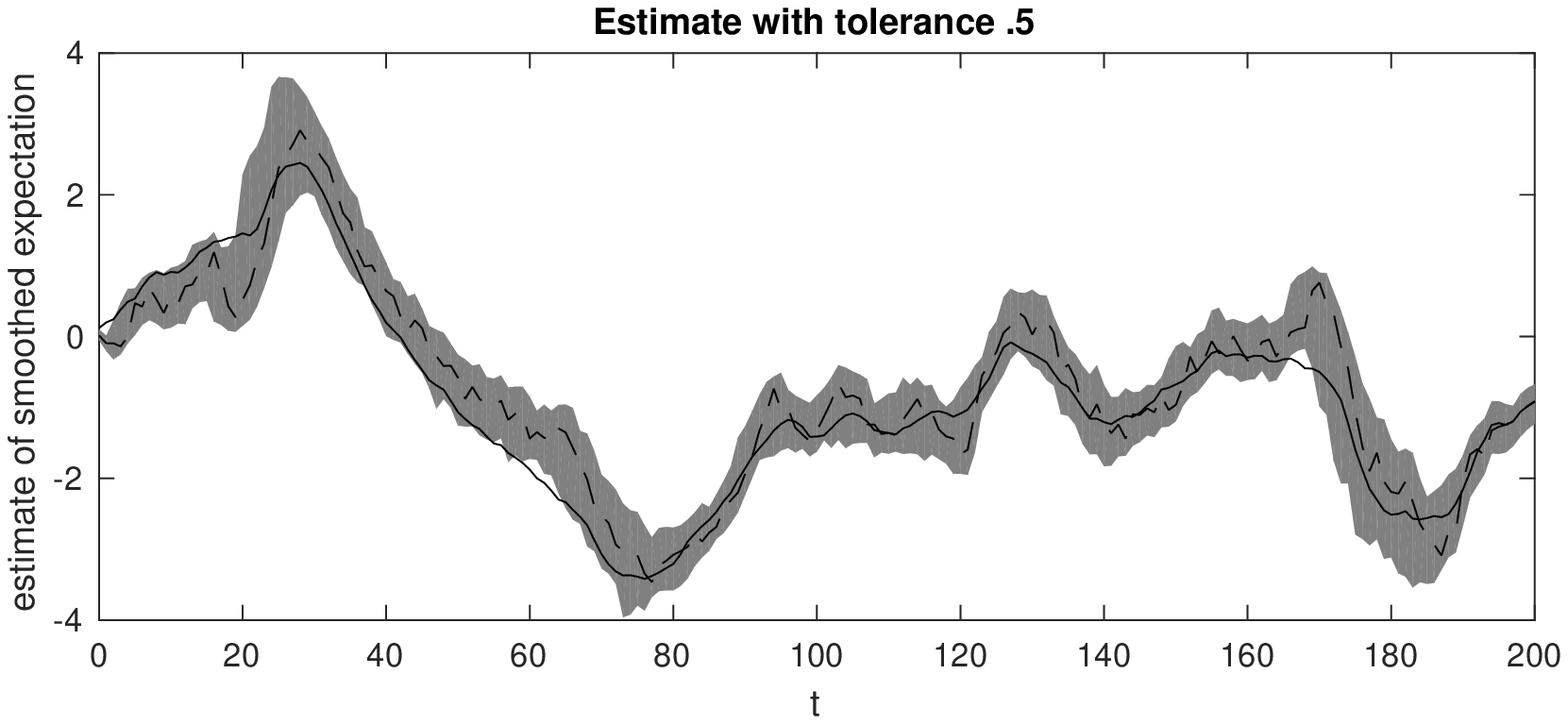}
\end{minipage}
\caption{Smoothed means for the linear Gaussian model. In each plot, the gray zone is the range of the particle-based estimates, while the black line indicates the exact smoothed values computed offline using the disturbance smoother. The dashed line is the Kalman version of our algorithm with the same tolerances as the particle-based method. When $\tol = 10^{-3}$ the Kalman filter-based estimates and the exact values are more or less indistinguishable.}
\label{fig:lg:ests}
\end{figure}

\begin{figure}
	\includegraphics[width = \linewidth]{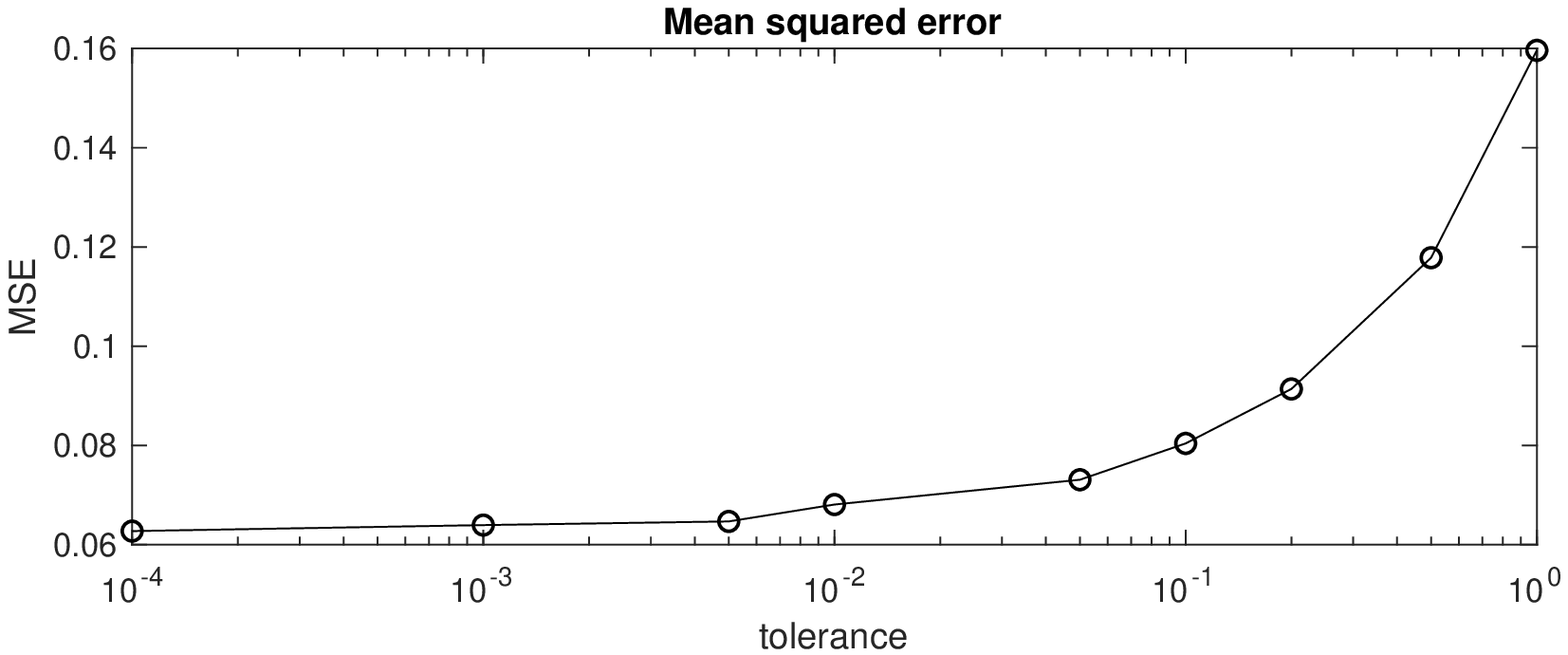} \vspace{.05cm}

	\includegraphics[width = \linewidth]{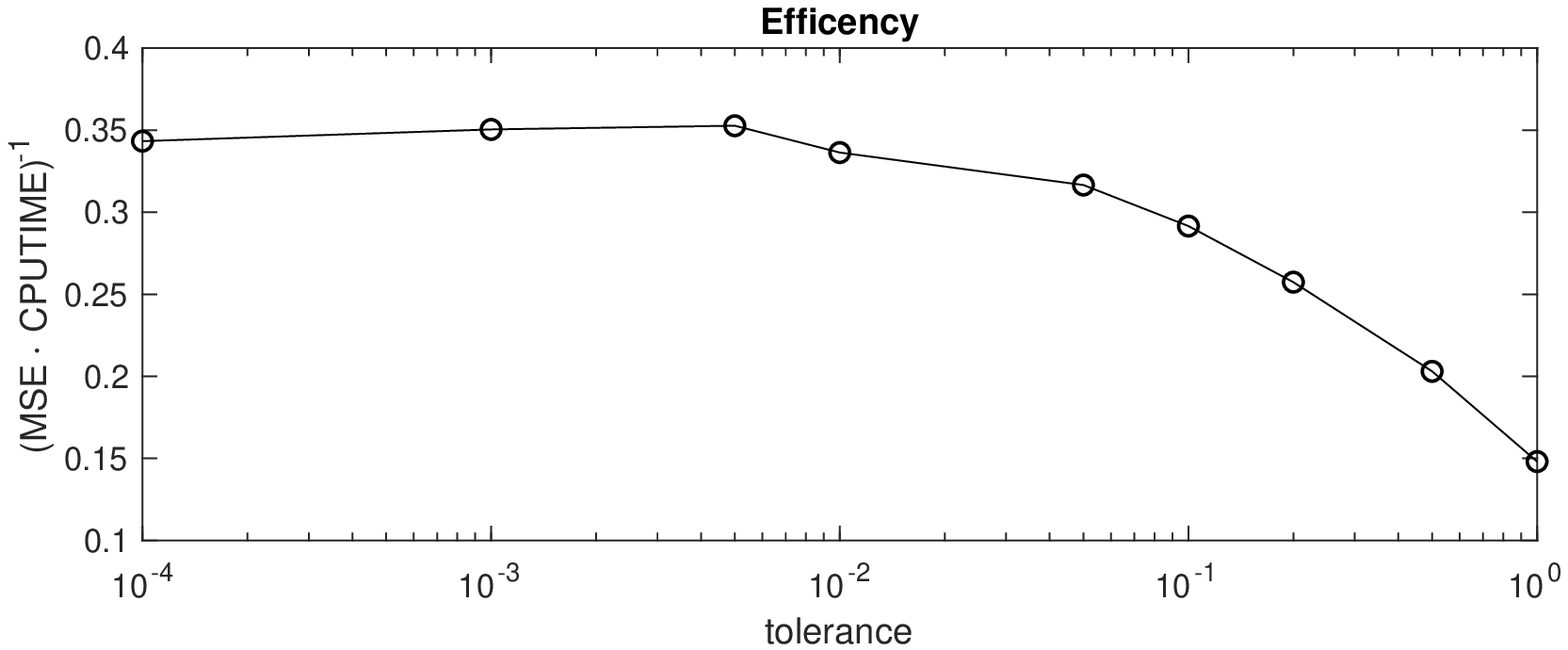}

	\caption{Top panel: Time-averaged MSE for different tolerances $\varepsilon$. MSE decreases monotonously with $\varepsilon$. Bottom panel: Efficiency, defined as the product of reciprocal MSE and reciprocal CPU runtime, against tolerance $\varepsilon$. Maximal efficiency is reached for $\tol = .5 \cdot 10^{-2}$.}
	\label{fig:lg:mse}
\end{figure}

First, the estimates produced by these two algorithms are compared to exact values computed offline using the disturbance smoother. The outcome is displayed in~\autoref{fig:lg:ests}, from which it is clear that the estimates improve with decreasing tolerance. For $\tol = .5$, the estimates exhibit clear fluctuations around the true values, while for $\tol = 10^{-3}$ the estimates follow closely the true quantities. The top panel of \autoref{fig:lg:mse} reports time-averaged mean squared errors (MSEs) (with respect to the exact posterior means delivered by the disturbance smoother) for different values of $\tol$. As expected, the MSE decreases monotonously with $\varepsilon$. However, since decreasing tolerance $\varepsilon$ comes at the cost of computational work, we also introduce a measure of \emph{efficiency}, defined as the product of reciprocal MSE and reciprocal CPU runtime. The bottom panel of \autoref{fig:lg:mse} reports efficiency for different values of $\tol$. Since the bias of the algorithm is more or less eliminated for all sufficiently small tolerances $\varepsilon$, there is an optimal value around $\varepsilon = .5 \times 10^{-3}$ for which the efficiency is maximal. 

\begin{figure}
	\includegraphics[width = \linewidth]{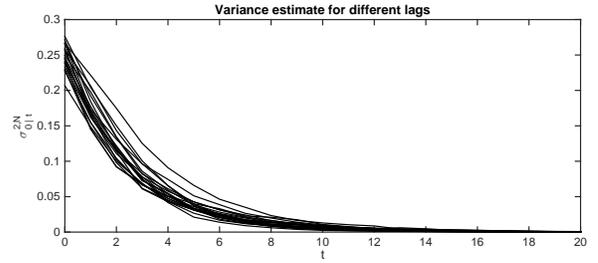}
	\caption{Linear Gaussian model: the initial variance criterion $\varlim[part]{0 \mid t}$ for different values of $t$.}
	\label{fig:lg:var}
\end{figure}

Finally, \autoref{fig:lg:var} displays the initial variance criterion $\varlim[part]{0 \mid t}$ for different values of $t$, and the plot is well in line with the exponentially decreasing bound provided by~\autoref{cor:bounded}. Finally, we study the truncation lags $s_{\tol}^{\N}(t) - s$ determined by the algorithm for different tolerances. The average lags across the $100$ runs are displayed in~\autoref{fig:lg:var}, from which it is evident that decreasing $\tol$ leads, in accordance with~\autoref{thm:activeEstimates}, on the average to larger lags. The slope in the end of the plot indicates non-truncation. 

\begin{figure}
	\includegraphics[width = \linewidth]{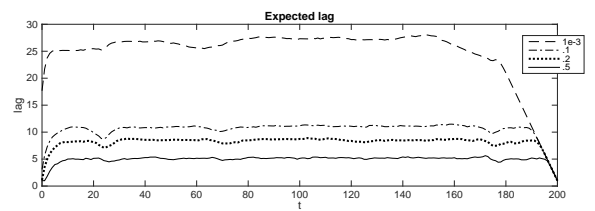}
	\caption{Linear Gaussian model: average lag against time $t$.}
	\label{fig:lg:lag}
\end{figure}

\subsection{Stochastic volatility model}
The second model of consideration is the stochastic volatility model 
\begin{align}
	X_{t+1} &= \phi X_t + \sigma U_{t+1}, \\
	Y_{t} &= \beta \exp(X_t/2) V_t,
\end{align}
where $\{U_t\}_{t \in \nset}$ and $\{V_t\}_{t \in \nset}$ are independent sequences of mutually independent standard Gaussian noise variables and $X_0 \sim \mathcal{N}(0, \sigma^2 /(1-\phi^2))$. In this nonlinear SSM, $\{Y_t\}_{t \in \nset}$ can be thought of as the log-returns of a stock while $\{X_t\}_{t \in \nset}$ can then be thought of as the unobserved log-volatility of the observed returns. 

As before, we simulate $201$ observations from the model indexed by $(\phi, \sigma, \beta) = (.98, \sqrt{.1}, \sqrt{.7})$. We employ the adaptive-lag smoother targeting again the mean of the marginal smoothing distribution, i.e., $\af{s} = \operatorname{id}$ for all $s \in \nset$. Algorithm~\ref{alg:afls} is run with the tolerances $\tol \in \{.5, .1, 10^{-3}\}$ and sample sizes $(\N, \K) = (400, 2)$. The algorithm is executed $200$ times for each value of $\tol$, and all runs are based on the same data input. Since exact computation is infeasible for this nonlinear model, we use, as reference, proxies for the true posterior means obtained as the averages of $10$ independent replicates of the full PaRIS (i.e., without stopping criterion) with $(\N, \K) = (2000, 2)$. 

\begin{figure}
\begin{minipage}{0.95\linewidth}
	\includegraphics[width = \linewidth]{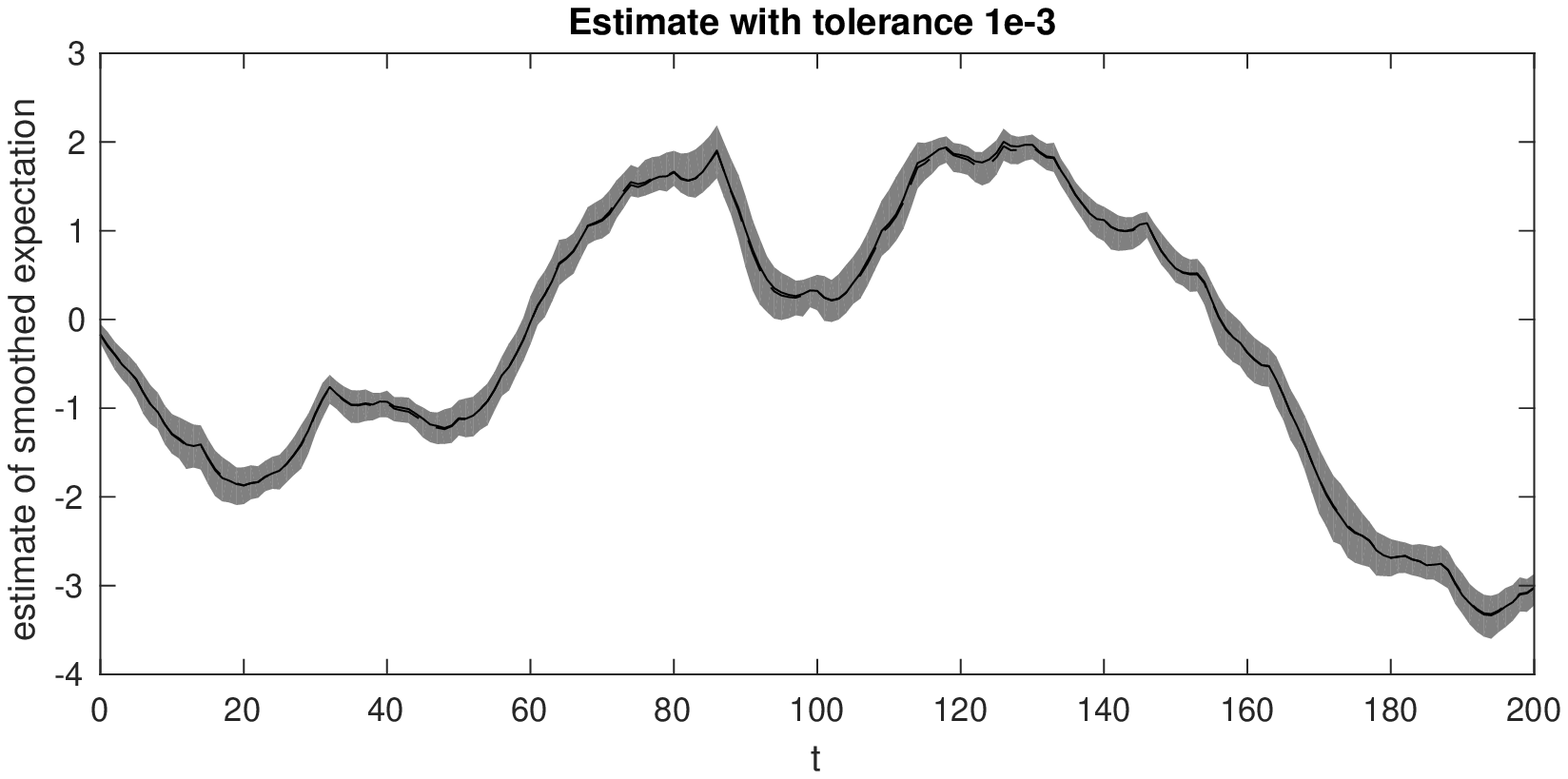}
\end{minipage}
	\begin{minipage}{0.95\linewidth}
	\includegraphics[width = \linewidth]{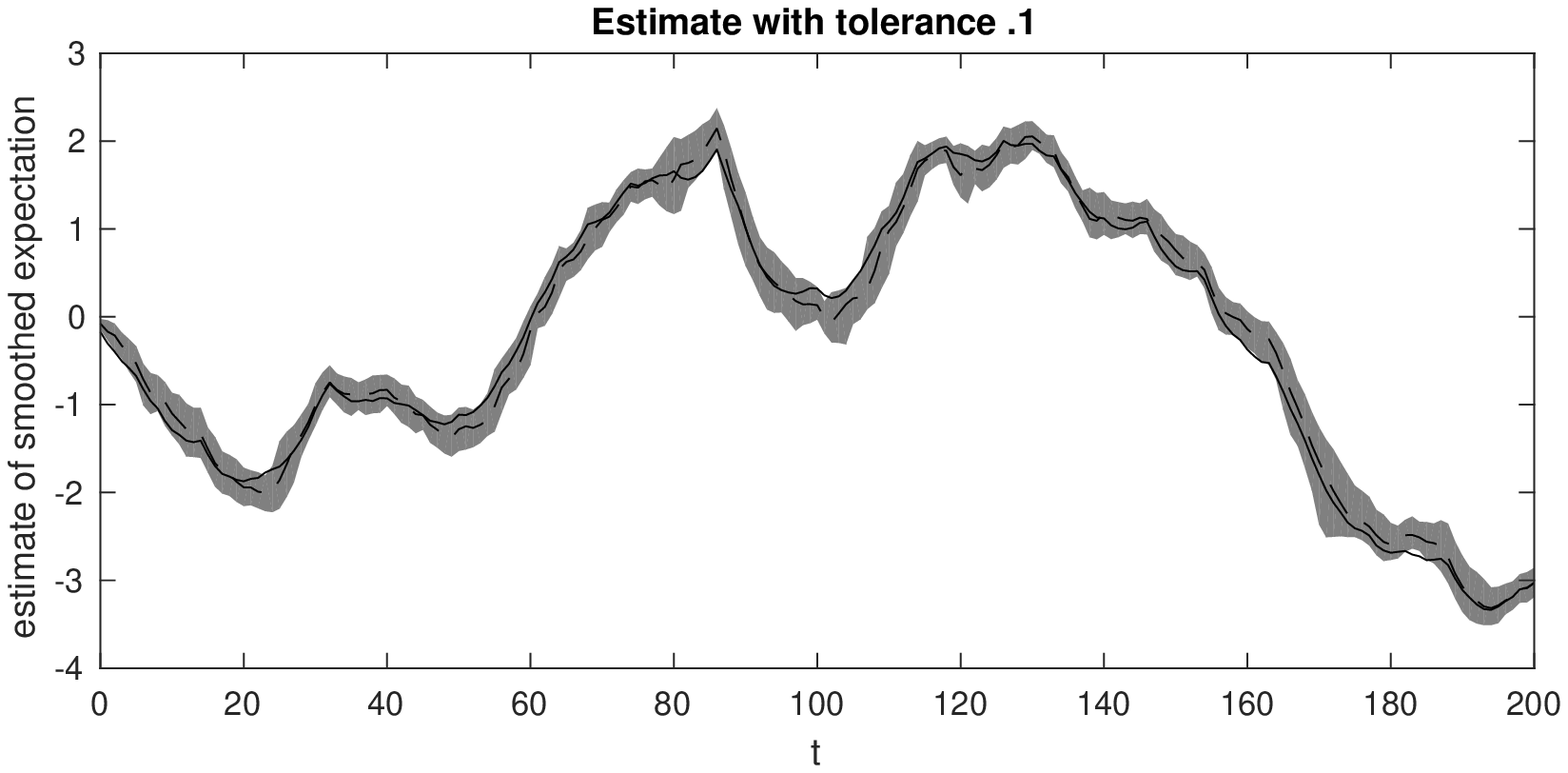}
\end{minipage}
\begin{minipage}{0.95\linewidth}
	\includegraphics[width = \linewidth]{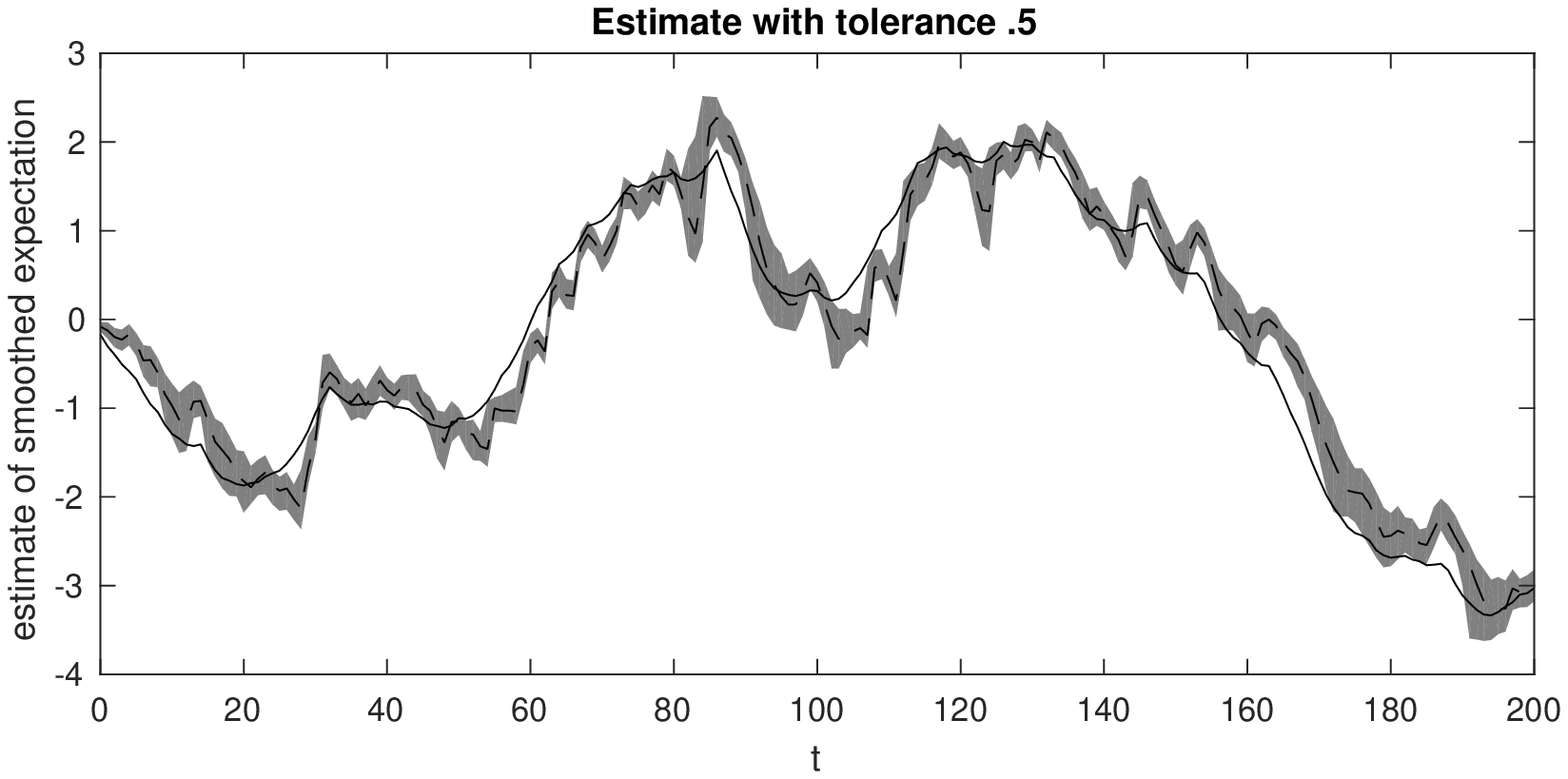}
\end{minipage}
\caption{Stochastic volatility model: smoothed means for different tolerances (cf. \autoref{fig:lg:ests}).}
\label{fig:sv:ests}
\end{figure}

The outcome is reported in~\autoref{fig:sv:ests}. For $\tol = .5$ the marginal smoothing estimates of the adaptive-lag smoother deviate significantly from the ground truth of the PaRIS algorithm for most time-steps. Decreasing the tolerance to $\tol = .1$ yields clearly improved---but still undesirably volatile---estimates. However, decreasing the tolerance even further to $\tol = 10^{-3}$ leads to a plot where the PaRIS estimates are firmly in the area of the estimates produced by the adaptive-lag smoother, and taking the mean over all the adaptive-lag estimates yields values indistinguishable from the estimates delivered by the PaRIS.

\begin{figure}
	\includegraphics[width = 0.95\linewidth]{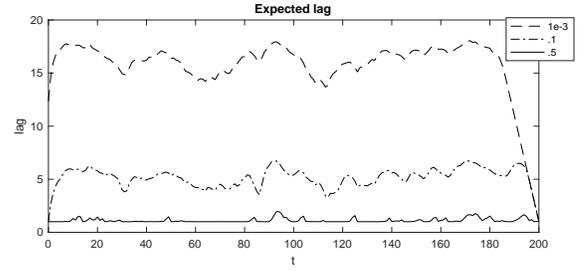}
	\caption{Stochastic volatility model: average lags for different tolerances.}
	\label{fig:sv:lag}
\end{figure}

Finally, \autoref{fig:sv:lag} provides the average lags at different time steps, and obviously the nonlinear components of the model leads to a higher degree of adaptation compared to the linear Gaussian model.

\subsection{Comparison with fixed-lag smoothing}
\begin{figure}
\includegraphics[width = \linewidth, trim = {0 0 0 0},clip]{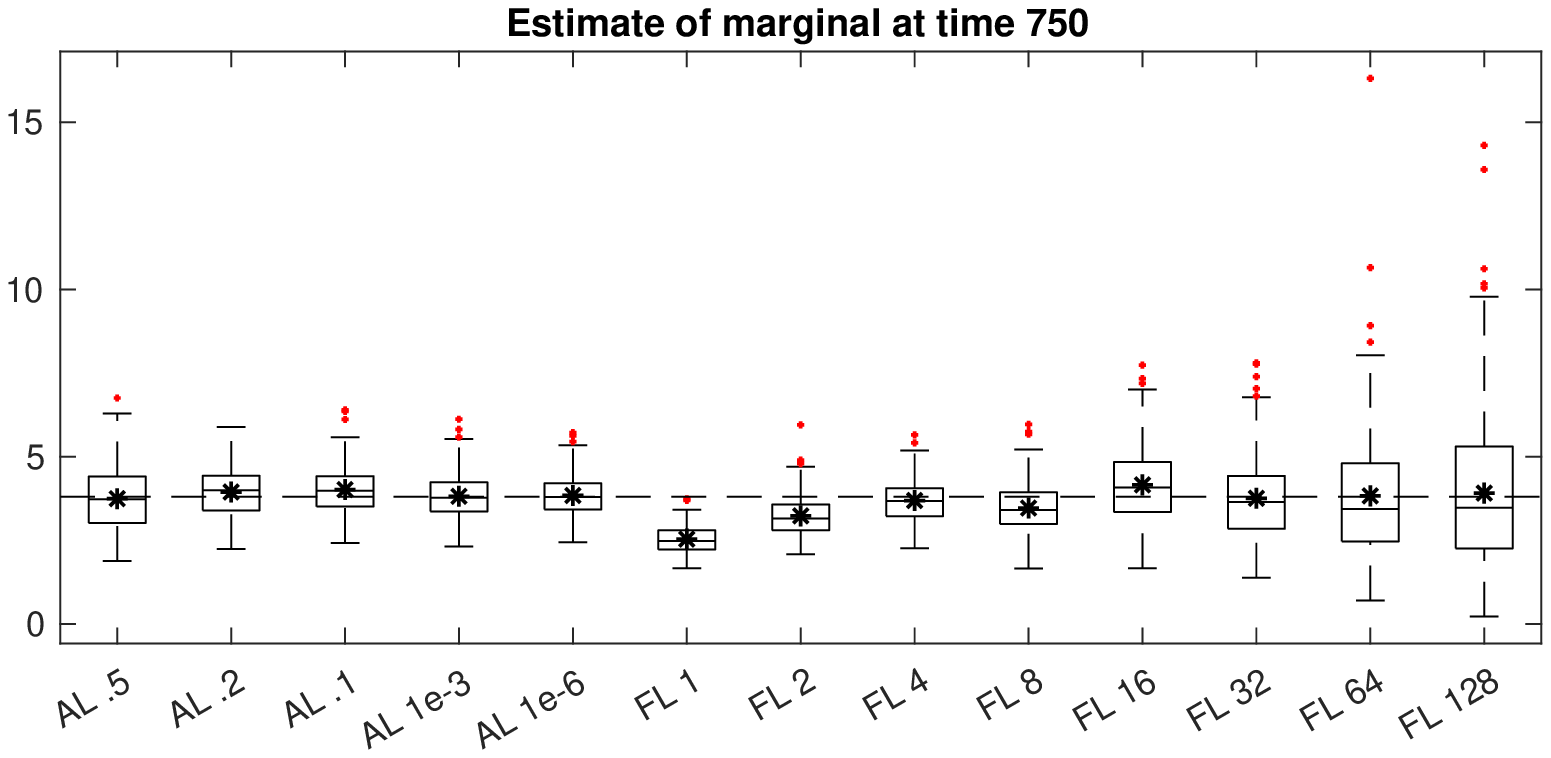}
\centerline{{\scriptsize (a) Linear Gaussian model}}

\vspace{.3 cm}\includegraphics[width = \linewidth, trim = {0 0 0 0},clip]{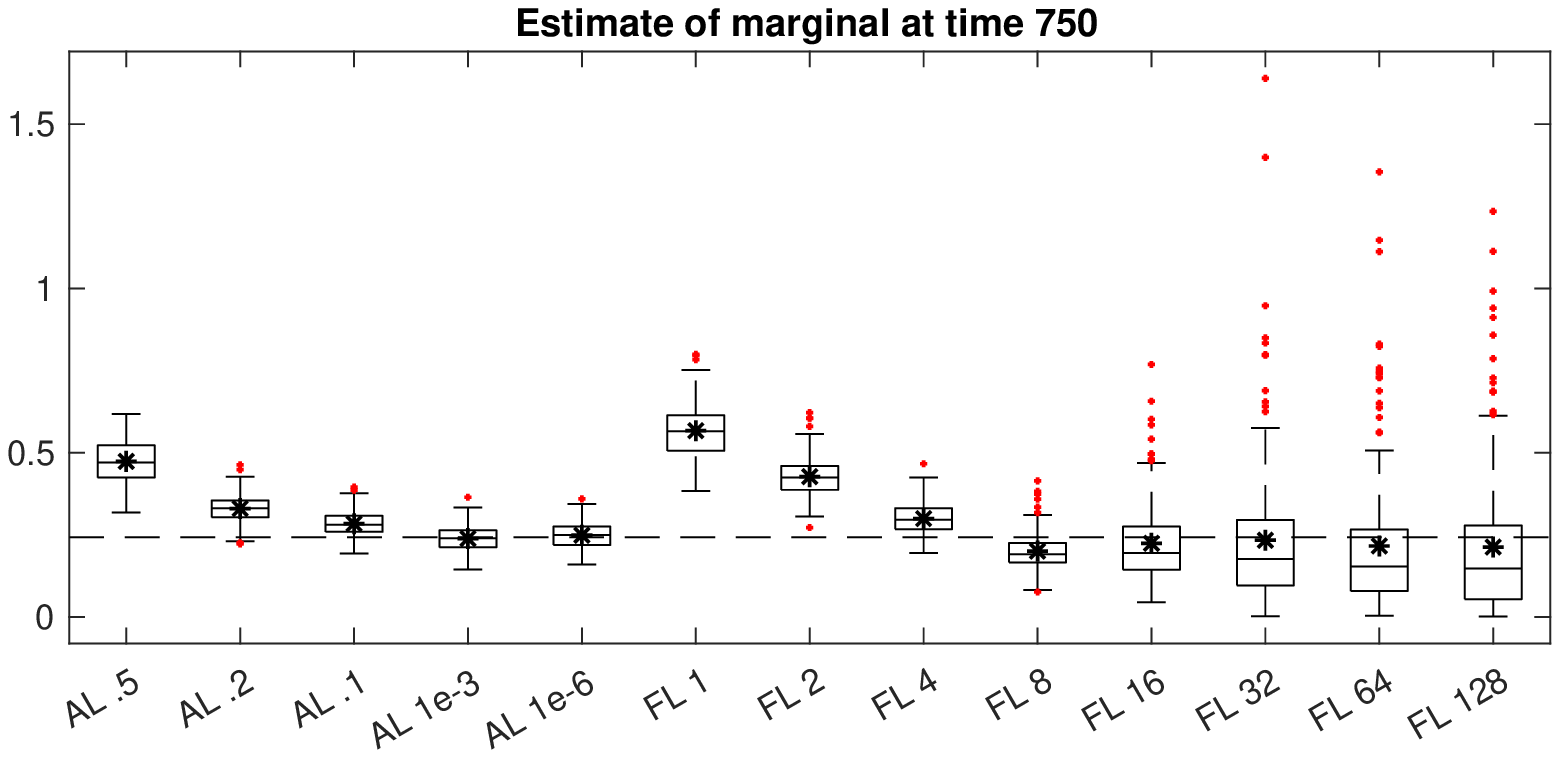}
\centerline{{\scriptsize(b) Stochastic volatility model}}

\caption{Comparison of the adaptive-lag smoother and the fixed-lag smoother for the linear Gaussian model (a) and stochastic volatility model (b). The black stars indicate the sample mean for each algorithm and the dashed line indicates a ground truth calculated exactly by the Kalman smoother in (a) and with the PaRIS algorithm in (b).}
\label{fig:FL:comp}
\end{figure}

We end the numerical illustrations with a comparison of the adaptive-lag smoother to the fixed-lag smoother. As mentioned earlier, in the fixed-lag smoother a lag has to be set a priori. The determination of the optimal lag requires the user to make a complex bias-variance tradeoff, governed by the mixing properties of the model as well as the coalescence properties of the particle paths. It is hence impossible to determine the optimal lag without first carrying through a preparatory simulation study. We simulate new data from both the linear Gaussian  and the stochastic volatility models using the same parameters as above and compare the adaptive-lag approach with $\tol \in \{.5, .2, .1, 10^{-3}, 10^{-6} \}$ and $(\N, \K) = (400,2)$ to the fixed-lag approach with lags $\Delta \in \{1, 2,4, 8, 16 ,32, 64, 128 \}$ and $\N = 400$. For both models we simulate $1001$ observations and aim at estimating $\E[X^2_s \mid Y_{0:1000}]$. We perform 200 independent runs of all algorithms. The results are displayed in~\autoref{fig:FL:comp}, where we have chosen $s = 750$ for illustration. From the plot it is clear that too small choices of the lag $\Delta$ lead to significant bias, while too large choices increase excessively the variance. The optimal choice of $\Delta$ can be seen to be somewhere around $\Delta = 8$ and $\Delta = 16$ for both models in this example. On the other hand, from the same plot it is clear that the tolerance $\tol$ does not obey the same bias-variance-tradeoff. When $\tol$ is simply small enough, the adaptive-lag smoother provides estimates having a variance being on par with that of the optimally tuned fixed-lag smoother. Moreover, importantly, decreasing excessively to $\tol = 10^{-6}$ does not, by the numerical stability of the PaRIS-type smoothing estimates, imply additional variance. 

\section{Conclusion}
\label{sec:conc}
In this paper we have presented a novel algorithm---an adaptive-lag marginal smoother---for online computation of marginal-smoothing expectations in general SSMs. We are not aware of any other algorithm in the literature solving satisfactorily this challenging problem. 


The proposed algorithm differs from the standard particle-based fixed-lag smoother \cite{kitagawa:sato:2001,olsson:cappe:douc:moulines:2006} (in our view, the only competitor of the proposed method) in essentially two ways: first, the estimates produced by the algorithm do not at all suffer from particle path degeneracy; second, the lag is designed adaptively by the algorithm via a variance criterion. Remarkably, since the PaRIS-type updates stabilize completely the support of the estimator as long as $\tilde{N} \geq 2$, the user does not risk afflicting the produced estimates with undesired variance by using an unnecessarily small tolerance $\tol$ and, consequently, an unnecessarily large average lag. This is the main advantage of the proposed algorithm over particle-based fixed-lag smoothing.



We remark that the proposed algorithm is, just like the PaRIS, \emph{function specific} in the sense that it is adjusted to the given sequence $\{ h_t \}_{t \in \mathbb{N}}$ of objective functions and outputs directly a sequence of estimated expectations rather than a sequence of weighted samples targeting the marginal smoothing distributions of interest. Thus, solving the problem \eqref{eq:tar} for \emph{different} objective function sequences requires different function-specific updates \eqref{eq:PaRIS-type:update} to be run in parallel. Still, nothing prevents these updates to be based on the same underlying particle filter and even the same backward draws $J_t^{(i, j)}$. 

Like all existing particle-based forward-filtering backward-smoothing-type algorithms, the proposed procedure relies on particle approximation of the backward kernels. Such approximation requires typically the SSM to be fully dominated. Even though many SSMs used in practice are indeed fully dominated, extending the methodology beyond this class of models is an important direction of research. Moreover, in our work we linearize the computational complexity of the backward sampling operation---the computational bottleneck of the algorithm---using the accept-reject approach proposed in \cite{douc:garivier:moulines:olsson:2010}. However, in practice, especially when the state space $\set{X}$ is high dimensional, the upper bound $\hkup$ may be very large, implying only a limited computational gain. In such cases, the adaptive-stopping approach suggested in \cite{taghavi:lindsten:svensson:schoen:2013} may come in useful. Finally, as in all backward-sampling-type smoothing algorithms, problems may occur for models and observation sequences for which the filter and smoothing distributions have significantly different support; we refer to \cite{ruiz:kappen:2017} for a discussion.    

When it concerns the theoretical developments, a result that is missing in the present paper is a rigorous theoretical analysis of the estimator's bias and the dependence of the bias on the tolerance $\tol$. Since the estimator is driven by ergodicity arguments, such an analysis requires control of the forgetting of the backward chain. However, this is still an open---and presumably very complex---problem, and the question is thus left as future research. 

Needless to say, the strong mixing assumptions (see \autoref{ass:strong:mixing}) driving the theoretical analysis of our estimator are restrictive. Still, these are standard in the literature, and only a few---technically involved---works analyze the time-uniform convergence of particle filters under weaker, verifiable assumptions that can be checked also for nonlinear SSMs with non-compact state space; see e.g. \cite{vanhandel:2009a,douc:moulines:olsson:2014}. When it concerns particle-based backward smoothing, \cite{jasra:2015} provides a stochastic stability analysis of the FFBSm algorithm under assumptions that point to applications with non-compact state spaces, and a similar approach may be applicable to our estimator.

The convergence results presented in \autoref{sec:theory} hold only asymptotically as $\N \to \infty$, and to furnish the estimator with error bounds for finite $\N$ would be appealing. When it concerns the PaRIS, it is indeed possible to derive an exponential concentration inequality for fixed $N$ (see \cite[Corollary~2]{olsson:westerborn:2014b}), and we thus expect such a Hoeffding-type bound to hold also for the estimator proposed in the present paper. However, to derive a \emph{time uniform} exponential concentration inequality for these algorithms appears to be considerably more complicated, as Hoeffding's inequality, which would be the first tool to try out in the proof, seems to be to crude for this purpose. Hence, we also have to leave this as future work.

\section*{Acknowledgment}

The authors would like to thank the anonymous referees for insightful comments that improved the presentation of the paper. The work of J.~Olsson is supported by the Swedish Research Council, Grant~2018-05230.

\bibliographystyle{plain}
\bibliography{biblio}

\appendix

\section{Proofs of theorems}
\label{sec:append}
We preface the proofs by some additional notation. 

For all $t \in \nset$, define the unnormalised transition kernels
\begin{align}
	\lk{t}(x_t, \rmd x_{t + 1}) \eqdef \md{t+1}(x_{t + 1}) \, \hk(x_t, \rmd x_{t + 1}),
\end{align}
with the convention that $\lk{s}\lk{t} \equiv \operatorname{id}$ when $s > t$. In addition, we let $\lk{-1}(x, \rmd x_0) \eqdef \md{0}(x_0) \, \delta_x(\rmd x_0)$. Moreover, we may express each joint smoothing distribution $\post{0:t \mid t}$ as $\post{0:t \mid t} = \post{t}\tstat{t}$, where we have defined the kernels
\begin{align}
	\tstat{t}(x_t, \rmd x_{0:t - 1}) \eqdef \begin{cases}
		\prod_{s = 0}^{t - 1} \bk{\post{s}}(x_{s + 1}, \rmd x_s) & \text{for } t \in \nsetpos, \\
		\operatorname{id} & \text{for } t = 0.
	\end{cases}
\end{align}
Notice that $\tstatfun{s \mid t}(x_s)$ in~\eqref{eq:tstat} can be expressed as
	$\tstatfun{s \mid t}(x_s) = \tstat{t}\af{s}(x_s)$.
Finally, define the operator   
\begin{align}
	\mathbf{D}_t : h \mapsto   
	\tstat{t}(\af{} - \post{0:t\mid t}\af{}), 
\end{align}
acting on the space of bounded measurable functions. 


\begin{proof}[Proof of \autoref{lem:var-conv}]
	By~\cite[Lemma~13]{olsson:westerborn:2014b} it holds that 
	\begin{equation}
		\sum_{i=1}^{\N}\frac{\wgt{t}{i}}{\wgtsum{t}}(\tstat[i]{s\mid t})^2 \convp \post{t}(\tstat{ t}^2\af{s}) + \sqc{t}, \label{eq:tsq:conv}
	\end{equation}
	where
	\begin{multline} \label{eq:def:eta}
		\sqc{t} \eqdef \\ \sum_{\ell = s}^{t-1} \K^{\ell - t} \frac{\post{\ell}\lk{\ell}\{\bk{\post{\ell}}(\tstat{ \ell}\af{s} - \tstat{ \ell + 1}\af{s})^2 \lk{\ell + 1} \cdots \lk{t-1} \1{\set{X}} \}}{\post{\ell}\lk{\ell} \cdots \lk{t-1}\1{\set{X}}}.
	\end{multline}
	In addition, from~\cite[Theorem~1]{olsson:westerborn:2014b} we get that
	\begin{equation}
		(\post[part]{s\mid t}\af{s})^2 \convp \post{t}^2(\tstat{ t}\af{s}),
	\end{equation}
	and combining the previous two limits yields, as $\N \rightarrow \infty$,  
	\begin{multline}
		\varlim[part]{s \mid t} = \sum_{i=1}^{\N} \frac{\wgt{t}{i}}{\wgtsum{t}}(\tstat[i]{s \mid t})^2 - \left( \sum_{i=1}^{\N} \frac{\wgt{t}{i}}{\wgtsum{t}} \tstat[i]{s \mid t} \right)^2 \\ 
		\convp \varlim{s \mid t} \eqdef \post{t}(\tstat{ t}^2 \af{s}) - \post{t}^2(\tstat{ t}\af{s}) + \sqc{t} \\
		=  \post{t}\{ (\tstat{ t} \af{s} - \post{t} \tstat{ t} \af{s})^2 \} + \sqc{t} .   
		\label{eq:varlim}
	\end{multline}
\end{proof}


\begin{proof}[Proof of \autoref{cor:bounded}]
	We begin by noticing that 
	\begin{equation}
		\post{t}\{ (\tstat{ t} \af{s} - \post{t} \tstat{ t} \af{s})^2 \} = \post{t}\left( \mathbf{D}_t^2 \af{s} \right). 
	\end{equation}
	Thus, by~\cite[Lemma~10]{douc:garivier:moulines:olsson:2010} it holds, for all $t \geq s$, 
	\begin{align}
		\supn{\mathbf{D}_t \af{s}} \leq \mr^{t-s} \hbdnt, \label{eq:Dtilbnd}
	\end{align}
	giving us the bound 
	\begin{align}
		\supn{ \post{t}(\mathbf{D}_t^2 \af{s})} \leq \mr^{2(t-s)} \hbdnt^2 
	\end{align}
        on the first term of~\eqref{eq:varlim}. 
	In order to bound the second term of~\eqref{eq:varlim}, i.e. $\sqc{t}$, we note that, \cite[Lemma~10]{douc:garivier:moulines:olsson:2010} yields that 
	$$
	\supn{\tstat{\ell}\af{s} - \tstat{\ell + 1}\af{s}} \leq 2 \mr^{\ell - s}\hbdnt.
	$$
        In addition, under \autoref{ass:strong:mixing}, for all $x \in \set{X}$,
	\begin{multline}
			\hklow \refM(\md{\ell + 2} \lk{\ell + 2} \cdots \lk{t-1}\1{\set{X}}) \\
			\leq \lk{\ell + 1}\cdots \lk{t-1}\1{\set{X}}(x) \leq \\
			\hkup \refM(\md{\ell + 2} \lk{\ell + 2} \cdots \lk{t-1} \1{\set{X}} ).
	\end{multline}
	Combining the previous two bounds allows $\sqc{t}$ to be bounded; indeed, proceed like
	\begin{multline}
		\frac{\post{\ell} \lk{\ell}\{ \bk{\post{\ell}}(\tstat{\ell}\af{s} - \tstat{\ell+1}\af{s})^2 \lk{\ell + 1}\cdots \lk{t-1} \1{\set{X}} \}}{\post{\ell}\lk{\ell}\cdots \lk{t-1}\1{\set{X}}} \\
		\leq 4 \mr^{2(\ell-s)} \hbdnt^2 \tfrac{ \hkup}{\hklow} = \mr^{2(t-s)}  \tfrac{4 \hbdnt^2}{(1 - \mr)}.
	\end{multline}
	Assuming now that $\K \mr^2 \neq 1$ we may bound $\sqc{t}$ using that
	\begin{align}
		\sqc{t} &\leq \frac{4 \hbdnt^2}{ ( 1 - \mr)} \sum_{\ell = s}^{t-1 } \K^{\ell - t} \mr^{2(\ell-s)} \\
		&= \frac{4 \hbdnt^2}{ ( 1 - \mr)} \K^{-t} \mr^{-2s} \sum_{\ell = s}^{t-1} (\K \mr^2)^{\ell} \label{eq:int:eqn}\\
		&= \frac{4 \hbdnt^2}{ (1 - \mr) ( 1 - \K \mr^2)}\left( \K^{-(t-s)} - \mr^{2(t-s)} \right).
	\end{align}
	On the other hand, if $\K \mr^{2} = 1$, \eqref{eq:int:eqn} yields 
	\begin{equation}
		\sqc{t} \leq 
		\tfrac{4 \hbdnt^2}{ ( 1 - \mr)} \K^{-(t-s)} (t - s).
	\end{equation}
	The previous argument may be summarised as 
	\begin{multline}
		\varlim{s \mid t} 
		\leq \hbdnt^2  
		 \begin{cases}
			c_1 \mr^{2(t-s)} + c_2 \K^{-(t-s)} & \text{if } \K \mr^{2} \neq 1, \\
			\mr^{2(t-s)} + c_3 (t-s) \K^{-(t-s)} & \text{if } \K \mr^{2} = 1,
		\end{cases}
	\end{multline}
	where 
		$$
		c_1 \eqdef 1 - \tfrac{4}{(1 - \mr)(1 - \K \mr^2)}, \ 
		c_2 \eqdef \tfrac{4}{(1 - \mr)(1 - \K \mr^2)}, \ 
		c_3 \eqdef \tfrac{4}{(1 - \mr)}.
	        $$
\end{proof}

\begin{proof}[Proof of \autoref{thm:convp}]
	Write 
	\begin{align}
		\postafl[part]{s \mid t}\af{s} &= \sum_{u = s+1}^{t} \1{\{s_{\tol}^{\N}(t) = u\}} \post[part]{s \mid u} \af{s} \\
		&= \sum_{u = s+1}^t \post[part]{s \mid u} \af{s} \1{\{ \varlim[part]{s \mid u} < \tol \}} \prod_{\ell = s + 1}^{u - 1} \1{\{ \varlim[part]{s \mid \ell} \geq \tol \}} \\ 
		&\hspace{30mm}+ \post[part]{s \mid t} \af{s} \prod_{\ell = s + 1}^t \1{\{ \varlim[part]{s \mid \ell} \geq \tol \}}. 
	\end{align}
	By \autoref{lem:var-conv} and \autoref{cor:bounded} it holds that $\varlim[part]{s \mid t} \convp \varlim{s \mid t}$ as $\N \to \infty$. 
	This yields, by Slutsky's theorem, as $\N \to \infty$, 
	\begin{multline}
	\postafl[part]{s \mid t}\af{s} \convp  \sum_{u = s+1}^t \post{s \mid u} \af{s} \1{\{ \varlim{s \mid u} < \tol \}} \prod_{\ell = s + 1}^{u - 1} \1{\{ \varlim{s \mid \ell} \geq \tol \}} \\ 
		+ \post{s \mid t} \af{s} \prod_{\ell = s + 1}^t \1{\{ \varlim{s \mid \ell} \geq \tol \}} 
		= \sum_{u = s+1}^{t} \1{\{ u = s_{\tol}(t) \}} \post{s \mid u} \af{s}. 
	\end{multline}
\end{proof}

\begin{proof}[Proof of~\autoref{thm:activeEstimates}]
	Note that 
	\begin{align}
		\act{t} = \sum_{s=0}^{t} \prod_{u = s}^{t} \1{\{\varlim{s \mid u} \geq \tol\}}\leq \sum_{s = 0}^{t} \1{\{\varlim{s \mid t} \geq \tol\}}.
	\end{align}
	Now, we consider separately the two cases $\K \mr^2 \neq 1$ and $\K \mr^2 = 1$. 

	First, assume that $\K \mr^2 \neq 1$.

	By \autoref{cor:bounded}, $\varlim{s \mid t} \geq \tol$ implies that $\hbdnt^2 c_1 \mr^{2(t-s)} + \hbdnt^2 c_2 \K^{-(t-s)} \geq \tol$. To proceed further, consider the two sub-cases (i): $\K^{-1} > \mr^{2}$ and (ii): $\K^{-1} < \mr^{2}$.
	\begin{itemize}
		\item[(i)] In this case, $c_1 < 0$; thus, if
		$$
			\hbdnt^2 c_1 \mr^{2(t-s)} + \hbdnt^2 c_2 \K^{-(t-s)} \geq \tol,
		$$
		then
		$$
			\hbdnt^2 c_2 \K^{-(t-s)} \geq \tol.
		$$
		Consequently, an upper limit on $(t - s)$ is given by
		\begin{align}
			(t - s) 
			\leq \tfrac{\log ( \tol / \{ \hbdnt^2 c_2 \})}{ \log \K^{-1}}.
		\end{align}
		Thus, $\act{t}$ may be bounded according to 
		\begin{multline}
			\act{t} \leq \sum_{s = 0}^{t}\1{\displaystyle\left\{ (t - s) \leq \tfrac{\log ( \tol / \{ \hbdnt^2 c_2 \})}{ \log \K^{-1}}\right\}} \\
			\leq \tfrac{\log ( \tol / \{ \hbdnt^2 c_2 \})}{ \log \K^{-1}}. 
		\end{multline}

	\item[(\emph{ii})] In this case, $c_2 < 0$; thus, as previously, 
	\begin{align}
		(t-s) \leq \tfrac{\log(\tol / \{\hbdnt^2 c_1\})}{\log \mr^{2}},
	\end{align}
	yielding the bound 
	\begin{align}
		\act{t} \leq \tfrac{\log( \tol / \{\hbdnt^2 c_1\})}{\log \mr^{2}}. 
	\end{align}
	\end{itemize}

	We now assume that $\K \mr^2 = 1$. In this case, $\varlim{s \mid t} \geq \tol$ implies that $\hbdnt^2 \K^{-(t-s)}\{1 + c_3 (t-s)\} \geq \tol$ as $c_3$ is always positive and $t \geq s$.  
	Thus,  
	\begin{align}
		(t-s) \leq \tfrac{\log(\tol / \{\hbdnt^2(1 + c_3)\})}{\log \K^{-1}}.
	\end{align}
	providing the bound 
	\begin{align}
		\act{t} \leq \tfrac{\log(\tol / \{\hbdnt^2(1 + c_3)\})}{\log \K^{-1}}.
	\end{align}

	Combining these results gives us the final bound
	\begin{align}
	 \act{t} \leq \tfrac{\log(\tol / \{\hbdnt^2 d(\mr, \K) \})}{\log (\mr^{2} \vee \K^{-1})}
	\end{align}
	for a constant $d(\mr, \K)$ depending on the constants $c_1$, $c_2$, and $c_3$ as well as the values of $\mr^2$ and $\K$. 
\end{proof}

\end{document}